\pdfoutput=1
\documentclass[prx,twocolumn,superscriptaddress,floatfix]{revtex4-2}
\usepackage[utf8]{inputenc}
\usepackage[english]{babel}
\usepackage[T1]{fontenc}
\usepackage{times}
\usepackage{stmaryrd}

\usepackage{lipsum}

\usepackage{amsmath, amssymb, amsfonts, amsthm, mathtools}
\usepackage{xspace, graphicx, relsize, bm, xcolor}

\usepackage{multirow}
\usepackage{physics}

\usepackage[linesnumbered,ruled,vlined]{algorithm2e}

\usepackage{enumitem}

\usepackage[colorlinks=true,linkcolor=teal,citecolor=teal,urlcolor=teal]{hyperref}

\usepackage{sepfootnotes}
\newendnotes{x}
\renewcommand\xnotesize\normalsize

\usepackage{url}
\usepackage{mleftright}

\usepackage{cleveref}

\newtheorem{theorem}{Theorem}
\newtheorem{observation}{Observation}
\newtheorem{lemma}{Lemma}
\newtheorem{corollary}{Corollary}

\newtheorem{definition}{Definition}
\newtheorem{proposition}{Proposition}

\newcommand{\jetext}[1]{#1}

\newcommand{\edit}[1]{\textcolor{black}{#1}}

\newcommand{\avg}[2]{\underset{#1}{\mathbb{E}}\left[#2\right]}

\newcommand{\Phik}{\mathbf{\Phi^{(k)}}}

\newcommand{\je}[1]{{\color{black}#1}}

\begin{document}

\title{Learning quantum functions using tensor cross interpolation}

\author{Seongwook Shin}

\affiliation{Dahlem Center for Complex Quantum Systems, Freie Universität Berlin, Berlin, Germany}

\affiliation{Department of Quantum Information Science and Engineering, Sejong University, Seoul 05006, Republic of Korea}

\author{Ryan Sweke}
\affiliation{African Institute for Mathematical Sciences, South Africa}
\affiliation{Department of Mathematical Sciences, Stellenbosch University, Stellenbosch 7600, South Africa}
\affiliation{National Institute for Theoretical and Computational Sciences (NITheCS), South Africa}
\affiliation{Dahlem Center for Complex Quantum Systems, Freie Universität Berlin, Berlin, Germany}

\author{Jens Eisert}
\affiliation{Dahlem Center for Complex Quantum Systems, Freie Universität Berlin, Berlin, Germany}
\affiliation{Helmholtz-Zentrum Berlin f{\"u}r Materialien und Energie, 14109 Berlin, Germany}
\begin{abstract}
We consider the problem of learning efficient classical representations of quantum functions, which provide an abstract representation of a wide variety of classically controlled quantum processes. Relevant examples include Hamiltonian simulators, which can be thought of as functions from Hamiltonian parameters to expectation values of a time-evolved state, or trained variational QML models, which are functions from circuit parameters to expectation values with respect to the output state of the circuit. We show that in many practically relevant cases, such quantum functions admit a tensor network representation, in terms of local function dictionaries that can be inferred from the parameter-dependent quantum gates. We then introduce and analyze an algorithm for learning efficient classical representations of such functions that we call function tensor cross interpolation. This algorithm builds on the recently developed technique of tensor cross interpolation, by lifting it to the setting of function dictionaries. To showcase the effectiveness of this learning algorithm, we provide a wide variety of numerical examples covering synthetic data, Hamiltonian simulators and quantum machine learning models.
\end{abstract}
 \maketitle

\section{Introduction}\label{sec:Intro}

The development of tools for gaining knowledge about unknown quantum systems has been a central theme in quantum information research. While 
early efforts largely focused on learning the full quantum state itself~\cite{anshu2024survey}, rapid progress in quantum-information techniques has broadened the range of objects of interest to include quantum channels \cite{PhysRevA.77.032322,AntonioChannels,RandomSequences}, 
quantum circuits \cite{CircuitLearning,LearningShallowCircuits2}, system Hamiltonians \cite{SampleEfficientHamiltonianLearning,HamiltonianLearning}, quantum resources \cite{MeasuringMagic,MeasuringEntanglement}, and more. This broad area is now in the process of being re-branded under the name of \textit{quantum learning theory}, and it continues to proliferate into more specialized questions and techniques. This body of work is highly relevant both from a conceptual as well as from a practical perspective, as one can only have predictive power if one has appropriately certified or learned the actual process or state at hand \cite{BenchmarkingReview,anshu2024survey}.

\jetext{One class of quantum processes} \jetext{that has so far received less attention} from a \jetext{learning-theoretic perspective} \jetext{is that of} \textit{classically controlled} quantum processes. \jetext{These are quantum processes that depend} on classical control parameters. As such processes are normally ultimately measured to extract classical information, one can think of these classically controlled quantum processes as \textit{quantum functions}---i.e., functions from classical control parameters to classical outputs, which are evaluated by the use of a quantum computational device. This approach naturally captures the functional dependence between control parameters and outputs of the device. \jetext{There exists a wide variety} of practically relevant examples of such functions, including:

\begin{enumerate}
\item \textbf{Hamiltonian dynamics:} Given the parameters of a Hamiltonian, evaluate the expectation value of some observable on a state evolved under the Hamiltonian.
\item \textbf{Inference in variational QML:} Given the parameters of a parameterized quantum circuit, evaluate the expectation value of an observable on the output state of the circuit.
\item\textbf{BQP:} Given a Boolean string $x$ and a quantum circuit, evaluate the expectation value of a computational basis measurement on the first qubit after applying the circuit to the state vector $|x\rangle$.

\end{enumerate}
We stress that the above list is not exhaustive, and in this work we call any function from classical information to classical information evaluated via the use of a quantum process a \textit{quantum function}. With this established, here we ask the question:

\begin{center}\textit{When and how can efficient classical representations of quantum functions be learned efficiently?}\end{center}

\jetext{The perspective we take is a marked departure from the usual approach to quantum dynamics. We do not seek to learn or classically simulate the evolution of a quantum state and then extract observables from this simulation. Instead, we take the functional dependence of measured quantities on classical control parameters as the object of interest in its own right. This distinction is central: The input--output map may admit a simple and efficiently learnable classical representation even when the underlying quantum-state evolution is difficult to simulate.}

Note, in particular, that we are specifically interested in obtaining \textit{efficient classical} implementations of quantum functions. \jetext{From the examples above, it should be clear} that whenever possible, this can be extremely useful. In particular, one can use some limited number of executions of a quantum process to generate a dataset sufficient for learning, and then subsequently use the learned classical function to evaluate the quantum function on future inputs, without having to 
run the quantum computational device. In this sense, our work can be viewed as a form of ``classical surrogation'', which has already been considered in the quantum machine-learning community~\cite{Schreiber2023surrogate}, but generalized to apply to \jetext{a broader class of classically controlled quantum processes}. We note 
that what we aim to achieve also has some similarities with classical shadows of quantum states~\cite{huang2020predicting}, which after some extraction of data from the unknown quantum state, allow one to classically evaluate (certain) expectation values with respect to the state. \jetext{Unlike classical shadows, however,} we are not focused purely on classical representations of quantum states, but of classically controlled quantum processes more broadly.

Of course, in general we do not expect to be able to obtain an efficient classical representation of \textit{any} controlled quantum process. As such, we focus here on characterizing a subset of meaningful and \jetext{physically well-motivated} quantum functions which do admit both efficient classical representations, and efficient classical learning algorithms. To do this, we leverage a novel kind of
\jetext{tensor cross interpolation} \cite{Yuriel2025learning}, a relatively new tool from the tensor network toolbox \cite{Orus-AnnPhys-2014, RevModPhys.93.045003,AreaReview}, both for representation and learning.

More specifically, in this work we do the following: We start in Section~\ref{sec:quantum-functions} by giving a rigorous definition of a quantum function, in the process specifying explicitly the class of classically controlled quantum processes that we consider. With this in hand, we then show in Section~\ref{ss:TN_repre_of_quantum} that every quantum function satisfying our definition admits a tensor network representation as the contraction of a coefficient tensor with a set of tensors representing local basis functions---what we refer to as \textit{function dictionaries}. Importantly, once the local function dictionaries are fixed, the function is fully specified by the coefficient tensor, and can be evaluated efficiently when this tensor admits a polynomial bond-dimension MPS representation. As such, learning the function reduces to learning (an efficient representation of) this coefficient tensor.

With this in mind, in Section~\ref{sec:FTQI} we propose and analyze an algorithm for learning functions which admit such tensor network representations. We call this algorithm \emph{function tensor cross interpolation}  (FTCI), given that it heavily exploits 
and builds upon the recently introduced technique of \emph{tensor cross interpolation}~\cite{Yuriel2025learning}. Once this algorithm has been established, we then proceed in Sections~\ref{sec:learning-q-functions} and~\ref{sec:conditions_for} to discuss the application of FTCI to learning quantum functions. In particular, we discuss complexities arising from the \textit{noisy} data resulting from actual quantum devices, and provide a rigorous characterization of the class of quantum functions whose coefficient tensor admits an efficient representation.

We then compare our FTCI algorithm for learning controlled quantum processes to related methods in Section~\ref{sec:comparison}, before providing a wide variety of numerical examples in Section~\ref{sec:numerics}. In particular, we study the FTCI algorithm on synthetic data, Hamiltonian simulation, and a QML model.

\section{Quantum functions and their tensor network representation}\label{sec:quantum-functions}

We start by providing the definition of a quantum function which is at the heart of our proposal. We stress that the following definition is not the most general definition one could consider\jetext{; however,} we will see that it is expressive enough to provide a natural way of capturing a large class of classically controlled quantum processes. To this end, let us denote \jetext{the} $n$-qubit Hilbert space \jetext{by} $\mathcal{H}^n$, the associated set of density matrices \jetext{by} $\mathcal{D}(\mathcal{H}^n)$, and the associated set of Hermitian matrices \jetext{by} $\text{Herm}(\mathcal{H}^n)$. We then adopt the following definition. 
\begin{definition}[Quantum function]\label{def:quantum_function}
Consider an external input (parameter) $x\in\mathbb{R}^d$ dependent quantum channel $\mathcal{C}_{\mathcal{G},\mathcal{P}}(x) : \mathcal{D}(\mathcal{H}^n) \rightarrow \mathcal{D}(\mathcal{H}^m)$, given as
\begin{equation}\label{eq:channel_structure}
    \mathcal{C_{G,P}}(x) = \prod_{k=1}^L \left(C_k\circ e^{-iG^{(k)}\phi^{(k)}(x)}\right)\circ C_0,
\end{equation} where $\circ$ denotes channel concatenation, $C_k$'s denote fixed 
quantum channels, $\mathcal{G}=\{G^{(k)}\}_k,\mathcal{P}=\{\phi^{(k)}\}_k$ denote multisets of generators, and pre-processing functions with $G^{(k)} \in \{ \text{Herm}(\mathcal{H}^l)~|~l\leq n\}$, and $\phi^{(k)} \in \{\phi : \mathbb{R}^d\rightarrow \mathbb{R}\}$. Given this, we define 
the \textit{quantum function} $f_Q(x;O,\mathcal{P,G})$ via
\begin{equation}\label{eq:quantum_function}
    f_Q(x;O,\mathcal{P,G}) := \Tr{\mathcal{C}_{\mathcal{G,P}}(x)[\rho] O},
\end{equation} where $O \in \text{Herm}(\mathcal{H}^m)$ is a fixed observable, and $\rho \in \mathcal{D}(\mathcal{H}^n) $ is a fixed initial density matrix.
\end{definition}

In other words, for the purposes of this manuscript, a quantum function is generated by a parametrized quantum channel, where the parameters control the unitary gates, while the gate and measurement settings are fixed. As already mentioned, this is certainly not the most general definition possible of a controlled quantum process, which maps classical control parameters to classical outputs. However, in the following section we will show that it is sufficient to capture a variety of practically relevant examples.

\subsection{Examples of quantum functions}

We now relate this abstract definition with natural processes \jetext{in the context of quantum technologies}. To make clear the relevance of Definition~\ref{def:quantum_function}, below we provide a (non-exhaustive) list of concrete motivating examples.\\

\noindent\textbf{Hamiltonian dynamics}: Given a family of local Hamiltonians $H(x)$ parametrized by parameters $x \in \mathbb{R}^d$, evaluate the expectation value of some observable $O$ with the evolved quantum state via the parametrized Hamiltonian for some time $t>0$. The function $f_Q :\mathbb{R}^d\times \mathbb{R} \rightarrow \mathbb{R}$ is then given by \begin{equation}
    f_Q(x,t) := \Tr[e^{-iH(x)t}\rho_0e^{iH(x)t}O].
\end{equation}

When we simulate Hamiltonian dynamics using a Trotterized circuit, it becomes a quantum function that has the (unitary) quantum channel
\begin{equation}
    \mathcal{C} := \biggl(\prod_{k=1}^{L}
    e^{-i H^{(k)}(x)\,t/K}\biggr)^{\!K},
\end{equation} where $K$ is the number of Trotterization steps, and \jetext{the $H^{(k)}$ are} the local terms comprising the original Hamiltonian. Here one can think of all fixed quantum channels \jetext{$C_i$} in Eq.~\eqref{eq:channel_structure} as identities.
\\

\noindent \textbf{QML models}: QML models that take classical inputs $x\in\mathbb{R}^d$, use a Hamiltonian data encoding strategy~\cite{Schuld2021effect}, and output a real number reflecting an expectation value. If one uses variational circuits for inference, variational parameters $\theta$ can also be thought of as the input of a quantum function.  
\smallskip

\noindent\textbf{BQP characteristic circuits}: Given a quantum circuit $U$ and a Boolean input $x\in\{0,1\}^d$, output the real number that is given by the probability of measuring the first qubit as $\ket{0}$. 
The quantum function in this context is given by 
\begin{equation}
    f_Q(x) = \Tr[U\ket{b}\bra{b}U^{\dagger}\ket{0}\bra{0}],
\end{equation} where $\ket{b}$ is a quantum state vector that represents the Boolean input. One accepts $x$ when $f_Q(x)$ is larger than certain threshold. This example corresponds to 
the case $L=d$, with 
    \begin{equation}
        C_k =\begin{cases} I,\quad k=[0,d-1],\\
    U,\quad k= d,
\end{cases}\end{equation}
$G^{(k)} = X^{(k)}/2$, and $\phi^{(k)}(x) = \pi x_k$.
\je{This list showcases that a large number of natural quantum processes and circuits can indeed be captured in this fashion.}

\subsection{Tensor network representation of quantum functions}\label{ss:TN_repre_of_quantum}

In this section, we show that quantum functions as per 
Definition~\ref{def:quantum_function} admit a natural tensor network representation
\cite{Orus-AnnPhys-2014, RevModPhys.93.045003,AreaReview}, yet one of a new kind tailored to the task at hand, and built on 
\jetext{quantum-inspired hypothesis classes} constructed from
\jetext{tensor-network-structured function dictionaries} \cite{QuantumInspired,MANDy,B01Goessmann,shin2025newperspectives}.
Concretely, to do this, given the set of generators $\mathcal{G} =\{G^{(k)}\}$ we denote by $\Lambda(G^{(k)})$ the set of eigenvalues of $G^{(k)}$.  With this, we can then define the \textit{frequency set} $\Omega^{(k)}$ via 
\begin{equation}\label{eq:frequency_set}
    \Omega^{(k)} := \{|\lambda^{(k)}_l-\lambda^{(k)}_m| \mid \lambda^{(k)}_l, \lambda^{(k)}_m \in \Lambda(G^{(k)})\} / \{0\}.
\end{equation}
Additionally, we define $D^{(k)} = |\Omega^{(k)}|$. We then have the following result.

\begin{proposition}[Tensor network representation of quantum functions]\label{prop:TN_repre}
Every quantum function can be represented as
\begin{align}\label{eq:TN_quantum_function}
    f_Q(x;O,&\mathcal{P,G}) = \left[\mathbf{C}(O,\{C_i\}_i)\right] \cdot \left[\bigotimes_{k=1}^{|\mathcal{G}|} \mathbf{\Phi}^{(k)}(G^{(k)},\phi^{(k)})(x)\right] \nonumber\\
    &=\vcenter{\hbox{\includegraphics[height=0.2\linewidth]{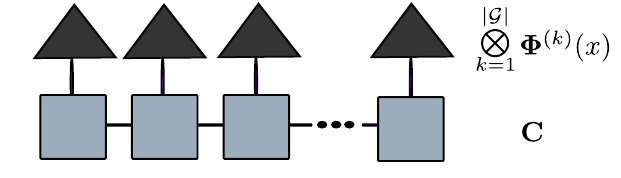}}},
\end{align}
where $\mathbf{C}(O,\{C_i\}_i)$ is a tensor with $|\mathcal{G}|$ outer legs, and
\begin{equation}\label{eq:canonical_basis}
    \mathbf{\Phi}^{(k)}(G^{(k)},\phi^{(k)})(x) := \frac{1}{\sqrt{D^{(k)}+1}}\begin{pmatrix}
        1\\
        \cos\left(\omega_1^{(k)}\phi^{(k)}(x)\right)\\
        \sin\left(\omega_1^{(k)}\phi^{(k)}(x)\right)\\
        \vdots\\
        \cos\left(\omega_{D^{(k)}}^{(k)}\phi^{(k)}(x)\right)\\
        \sin\left(\omega_{D^{(k)}}^{(k)}\phi^{(k)}(x)\right)\\
    \end{pmatrix}
\end{equation}
are rank-one tensors (vectors) of dimension $2D^{(k)}+1$, with $\omega^{(k)}_i \in \Omega^{(k)}$ as defined in Eq.~\eqref{eq:frequency_set}.
\end{proposition}

To avoid notational clutter, we omit the dependence on $O,\mathcal{G},\mathcal{P}$ from here on. We note also that this is a generalized version of the result from Ref.~\cite{shin2024dequantizing}, and the full proof can be found in Appendix~\ref{app:proof_of_prop1}. Additionally, it may also be viewed as a reiteration of the well-known fact that parametrized circuits correspond to truncated Fourier series~\cite{Schuld2021effect}. However, we emphasize again the key observation made here (also made in Ref.~\cite{sweke2025kernelbased}), which is that quantum functions admit a natural decomposition in terms of basis functions with a \textit{tensor product} structure. Indeed, this structure will soon play a crucial role. Before proceeding, we make the following remarks:
\begin{enumerate}
    \item \textbf{The coefficient tensor $\mathbf{C}$}: This is determined solely by the input-\textit{independent} parts of the system. An explicit formula for it is quite cumbersome, and we do not need it for the main results; therefore, we defer it to the appendix. Also, we note that obtaining $\mathbf{C}$ from the description of the quantum function requires the multiplication of exponentially large (with respect to the system size) matrices \jetext{(see Appendix~\ref{app:proof_of_prop1}),} which is not efficient in general. Tensor $\mathbf{C}$ can have many different structures, but in this work we set it to be in a \emph{matrix product state} (MPS) 
    form.

    \item \textbf{The basis-function dictionaries $\mathbf{\Phi}^{(k)}(x)$}: They are $2D^{(k)}+1$ dimensional vectors that contain basis functions. They are normalized in the sense that $\left[\mathbf{\Phi}^{(k)}(x)\right]^\top \cdot \mathbf{\Phi}^{(k)}(x) = 1$ for all $x$. The frequency sets $\{\omega^{(k)}\}$ are obtained from all possible differences among eigenvalues of the corresponding generators. For a generator $G^{(k)} \in \text{Herm}(\mathcal{H}^l)$, the number of frequencies satisfies 
    \begin{equation}
    D^{(k)} \leq \begin{pmatrix} 2^l \\ 2 \end{pmatrix}, 
    \end{equation}
    and the maximum can be achieved using the so-called Golomb encoding Hamiltonian introduced in Ref.~\cite{Peters2023generalization}. We refer to the dimensions of the basis function dictionaries as \textit{local dimensions}. 

    \item \textbf{Non-uniqueness}: The same function $f_Q$ can have many different TN representations. One can choose different function dictionaries $\{\mathbf{\Phi}'^{(k)}(x)\}_k$ as long as they contain all basis functions used by the quantum function. In that case, there exists a new coefficient tensor $\mathbf{C}'$ that yields the same quantum 
    function
    \begin{equation}
        f_Q(x) = \mathbf{C} \cdot \bigotimes_{k=1}^L \mathbf{\Phi}^{(k)}(x)
        = \mathbf{C}' \cdot \bigotimes_{k=1}^{L'} \mathbf{\Phi}'^{(k)}(x).
    \end{equation}
    Note that the number of function dictionaries, as well as their local dimensions, may differ. Because of this non-uniqueness, we call the TN representation obtained using the function dictionaries in Eq.~\eqref{eq:canonical_basis} the canonical form. One useful non-canonical form is to combine all basis functions that depend on the same pre-processing function $\phi$. 
    \jetext{This makes the length} of the TN representation $L' = |\mathcal{P}|$, instead of $|\mathcal{G}|$, see 
    Fig.~\ref{fig:Nonuniquness}\jetext{.}

    \item \textbf{Possible efficient representation}: The total number of basis functions can scale as $O(D_{\max}^{|\mathcal{G}|})$, where 
    \begin{equation}
    D_{\max} := \underset{k}{\max}[ D^{(k)}], 
    \end{equation}
    which is exponential with respect to the number of input-dependent 
    unitaries. This renders the representation of quantum functions inefficient in 
    the naive linear sum representation. However, with the TN representation, one can 
    see that whenever the coefficient MPS allows efficient representation (i.e., having $O(\mathrm{poly}(|\mathcal{G}|))$ bond dimensions), the quantum function can be stored (and evaluated) efficiently. This tells us that an exponentially large number of basis functions themselves is not a hurdle for efficient representation, but the limiting factor is the structure of the coefficient tensor.
\end{enumerate}
Such a tensor network decomposition is natural in situations with short-ranged correlations in a precise sense.

\begin{figure*}
    \centering
    \includegraphics[width=1.0\linewidth]{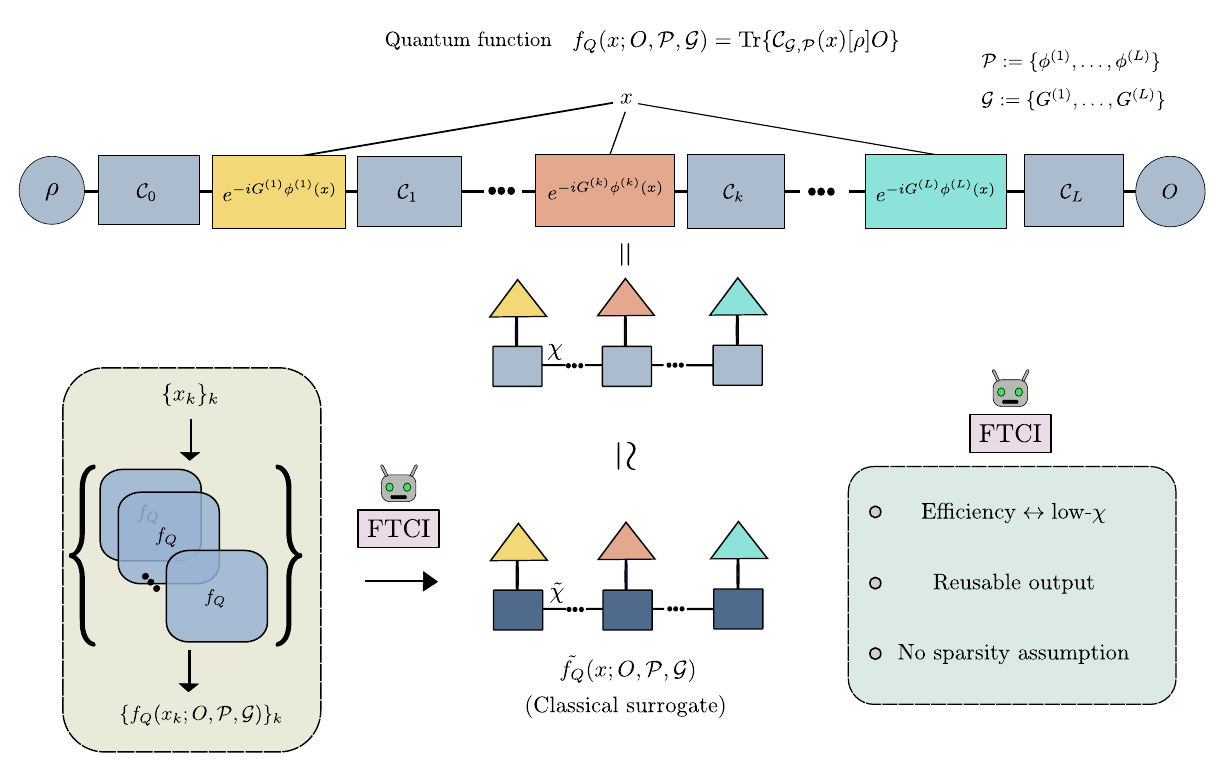}
    \caption{Overview of the approach taken in this work: We consider a quantum system that is controlled by an external classical input $x\in\mathbb{R}^d$ through input-dependent unitaries generated by a multiset of Hermitian generators $\mathcal{G}$ and a multiset of pre-processing functions $\mathcal{P}$ that map inputs to rotation angles. We assume that $\mathcal{G}$ is known \emph{a priori}. The quantum system outputs the expectation value of a fixed observable $O$, \jetext{which depends on $x$; we refer} to the resulting input--output map as a \emph{quantum function}. We identify these quantum functions \jetext{as a contraction between function-dictionary tensors and a coefficient tensor}. Then, we propose a general function-learning algorithm \textit{function tensor cross interpolation} (FTCI), that is \jetext{a variant of the tensor cross interpolation algorithm} \jetext{to reconstruct the coefficient tensor in matrix product form} given function basis dictionaries. We propose using such FTCI to learn quantum functions \jetext{for large system sizes}, with sample complexity scaling \jetext{polynomially in $|\mathcal{G}|$ and $\chi$}, where $\chi$ is the maximum bond dimension of the coefficient MPS. After learning, we obtain an efficient tensor network representation of the quantum function that is efficient to implement on a classical machine. The efficiency guarantee comes from the compressibility of \jetext{the coefficient tensor into an efficient MPS}, and it does not require a sparsity assumption, unlike many other function-learning algorithms. }
    \label{fig:overview}
\end{figure*}

\subsection{Example: TN representation of a simple quantum function}\label{ss:example_encoding_first}

\jetext{To provide intuition, we provide} here an example of a simple quantum function that takes input $x \in \mathbb{R}^n$ and outputs a real number in $[-1,1]$. We construct this function considering an $n$-qubit state vector $|0\rangle^{\otimes n}$\jetext{,} acted on by an input-dependent unitary channel,
\begin{align}
    U(x) = W\bigotimes_{k=1}^n e^{-i\frac{Z^{(k)}}{2}\phi^{(k)}(x)}H^{(k)},
\end{align}where $Z^{(k)}$ and $H^{(k)}$ are Pauli-Z and Hadamard operators acting on the $k$'th qubit, and $W$ is an input-independent unitary. To obtain the output, we measure some observable $O$. Putting it together, we have
\begin{align}
f_Q(x) &= \mathrm{Tr}\left(OU(x)\ketbra{0}{0}^{\otimes n}U^\dagger(x)\right), \nonumber\\
&=\bra{0}^{\otimes n}U^{\dagger}(x)OU(x)\ket{0}^{\otimes n},
\end{align}
and in the language of Definition~\ref{def:quantum_function}, 
we have,
\begin{align}
    \rho &= \ketbra{0}{0}^{\otimes n}, \\
    \mathcal{G} &= \{Z^{(k)}/2~|~k \in [n]\}, \\
    \mathcal{P} &= \{\phi^{(k)} ~|~k \in [n]\}, \\
    C_0 &= H^{\otimes n}(\cdot) H^{\otimes n},~C_{1\leq k \leq n-1} = I(\cdot)I, ~C_n = W.
\end{align}
This function can now be written in length-$n$ TN form with function dictionaries
\begin{align}
    \mathbf{\Phi}^{(k)}(x) = \frac{1}{\sqrt{2}}\begin{pmatrix}
        1\\\cos \phi^{(k)}(x) \\ \sin \phi^{(k)}(x) 
    \end{pmatrix}.
\end{align} Interestingly, for this special structure, the coefficient tensor $\mathbf{C}$ \jetext{is given by}~\cite{shin2024dequantizing}
\begin{equation}\label{eq:coefficient_special}
    \mathbf{C}_{i_1, \ldots , i_n} = \frac{1}{2^{n/2}}\Tr{P_{i_1,\ldots , i_n} W^{\dagger}OW},\quad i_k \in \{0,1,2\},
\end{equation} where $P_{i_1, \ldots , i_n}$ is 
a Pauli string with
\begin{equation}
    P_{0} = I,~P_{1} = X,~P_{2} = Y.
\end{equation}
In other words, the entries of $\mathbf{C}$ are the coordinates of $W^{\dagger}OW$ in the Hilbert--Schmidt-orthonormal Pauli basis restricted locally to $I$, $X$, and $Y$.
By choosing different pre-processing functions,  one can generate many different functions. One notable case is when one chooses
\begin{align}
    \phi^{(k)}(x) = \pi x_k,
\end{align} with $x \in \{0,1\}^n$, and the observable $O = Z^{(1)}$.
Now this gives us a length-$n$ TN representation with function dictionaries,
\begin{align}
    \mathbf{\Phi}^{(k)}(x) = \frac{1}{\sqrt{2}}\begin{pmatrix}
        1\\\cos \pi x_k \\ \sin \pi x_k
    \end{pmatrix} = \frac{1}{\sqrt{2}}\begin{pmatrix}
        1\\(-1)^{x_k}\\0
    \end{pmatrix},
\end{align} whose entries are functions on $\{0,1\}$ that can also be used to solve decision problems. In this case, the basis functions reduce to the Fourier basis for Boolean functions, while the coefficient tensor contains the Hilbert-Schmidt coordinates of $W^{\dagger}Z^{(1)}W$ restricted locally to $I$ and $X$.

\begin{figure}
    \centering
    \includegraphics[width=1.0\linewidth]{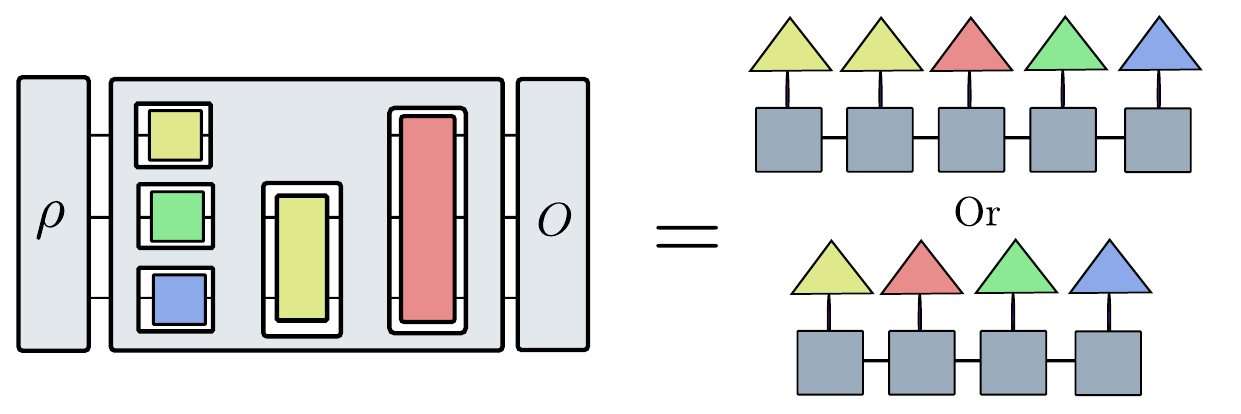}
    \caption{Non-uniqueness of TN representation. Colored boxes are input-dependent unitaries, where all generators could be different. Colored triangles represent function dictionaries. We used the same color when they depend on the same scalar, $\phi^{(k)}(x)$. The above one is the canonical representation where we use $|\mathcal{G}|$ function dictionaries, while the lower one uses only $|\mathcal{P}|\leq|\mathcal{G}|$ of them. Note that we could have more representations than those depicted}
    \label{fig:Nonuniquness}
\end{figure}

\section{Function tensor cross interpolation}\label{sec:FTQI}

Given black box access to a quantum function, our objective is to learn an efficient classical representation of this function, so that in order to evaluate the function we no longer need to access the quantum circuit or system after the initial learning phase. In the previous section, we saw that characterizing a quantum function can be reduced to identifying its coefficient tensor. Moreover, the TN structure makes clear that if this coefficient tensor admits an MPS representation with small bond dimension, then the quantum function itself has an efficient classical representation. Here we present an algorithm to learn the coefficient tensor (and hence the entire function, given the basis-function dictionaries) whenever the coefficient tensor admits an efficient MPS representation.

To do this,  we introduce \emph{function tensor cross interpolation} (FTCI) which can be used to learn tensor network representations of functions which admit such a structure. Specifically, given local basis functions or function dictionaries, it approximates the coefficient tensor of the target function in the form of MPS by using \textit{tensor cross interpolation} (TCI)~\cite{Yuriel2025learning} as a subroutine. Given this, we briefly introduce the TCI algorithm and then move on to \jetext{our presentation} of FTCI. 

\subsection{Tensor cross interpolation}

We first recall how tensor cross interpolation builds a compact MPS from selected tensor entries.
For this, consider an $L$-leg tensor $\mathbf{T}$ with local dimensions $d_k$ satisfying $d_{\mathrm{\mathrm{max}}}:=\underset{k}{\max}~{d_k}$. When representing such an arbitrary tensor $\mathbf{T}$ as an MPS, the maximum bond dimension of $\mathbf{T}$, which we denote with $\chi_\mathbf{T}$ could be $O(d_{\mathrm{\mathrm{max}}}^{L/2})$. However, \jetext{some such tensors can be approximated} well by an MPS $\hat{\mathbf{T}}$ with much smaller bond dimension $\chi \ll\chi_\mathbf{T}$. This low-rank (-bond dimension) approximation/compression can be done by repeatedly applying SVD on the given tensor, and achieves the error 
\begin{equation}
\varepsilon_{\chi}^* := \norm{\bf{T} - \hat{\mathbf{T}}}_F , 
\end{equation}
and this is the best one can hope for low-rank approximation. To accomplish this compression, one needs to access all $O(d_{\rm max}^L)$ elements, which is \jetext{scaling exponentially} with the tensor size.

However, \jetext{there exists a suboptimal compression algorithm} using only access to $O(d^2_{\mathrm{\mathrm{max}}}\chi^2L)$ elements of $\bf{T}$, called \textit{tensor cross interpolation} (TCI)~\cite{waintal2026competequantumcomputerslecture,Yuriel2025learning}. This works by adaptively choosing entries from $\bf{T}$ to query, and compressing relevant matrices using cross interpolation, instead of SVD and truncating. Of course, one loses accuracy as a tradeoff of improved sample and time complexity. However, perhaps surprisingly this accuracy cost is not \jetext{as bad as one might fear}---in particular, the accuracy of TCI is only polynomially larger than $\varepsilon_{\chi}^*$, and grows with the power of $O(\log(L))$~\cite{qin2022error}. 

We note that one can also derive error bounds for the noisy case where one queries from $\bf{T} + \mathbf{\mathcal{E}}$, instead of $\mathbf{T}$. In particular, this error bound shows that noisy evaluations introduce only quasi-polynomially scaling additive error with respect to $L$~\cite{qin2022error}.

In summary, TCI is an algorithm for learning, compressing, and approximating a tensor using an MPS ansatz 
with designated maximum bond dimension $\chi$. It is suboptimal in the sense of accuracy, but is exponentially more efficient in time and sample resources, when $\chi$ is small enough. The additional error compared to the optimal error method is (quasi-) polynomially related.
For a detailed explanation and algorithm, we refer to the cited literature. 

\subsection{Function tensor cross interpolation}

In this subsection, we take the step towards \emph{function tensor cross interpolation} (FTCI). Specifically,
the objective here is not to learn a tensor as such, but rather to learn a function $f: \mathbb{R}^d \rightarrow \mathbb{R}$ with some assumed tensor network structure.  As the algorithm we present strongly exploits tensor cross interpolation as a subroutine, we therefore refer to it as \textit{function tensor cross interpolation}.

\subsubsection{Assumptions}\label{ss:assumptions}

In order to exploit tensor cross interpolation, we require the target function to satisfy a number of assumptions, \jetext{which ensure that the function} admits a specific tensor network structure. While these assumptions are motivated by the tensor network representation satisfied by quantum functions, we phrase them here in a more general way. More specifically, we naturally 
assume the following:

\begin{enumerate}
    \item We have query access to $f:\mathbb{R}^d\rightarrow \mathbb{R}$.
    \item The target function $f$
    is a finite linear combination of some efficiently computable functions---i.e.,
    \begin{equation}
        f(x) = \sum_{i=1}^Dc_ib_i(x).
    \end{equation} Neither the coefficients $\{c_i\}$ nor the functions $\{b_i\}$ are necessarily known. 
    \item For some $L\leq d$, we are given $L$ sets $B^{(k)}$ for $k\in [L]$, where
    \begin{equation}
    B^{(k)}= \{B^{(k)}_{i_k}(x)\,|\, i_k\in[Q^{(k)}]\}
    \end{equation}
    is a size $Q^{(k)}$ set of functions from $\mathbb{R}^d\rightarrow\mathbb{R}$. We then 
    denote by $\mathbf{B}^{(k)}(x)$ the length $Q^{(k)}$ column 
    vector with 
    $B^{(k)}_{i_k}(x)$ in the $i_k$'th row. With this notation, we make the 
    additional assumption that the span of the functions which form the rows of the vector $\bigotimes_{k=1}^L\mathbf{B}^{(k)}(x)$ includes the target function $f$.
    \item There exists a size-$L$ partition of $[d]$, namely a size-$L$ set $\{J_1,\ldots,J_L\}$ of subsets of $[d]$, satisfying
    \begin{enumerate}
    \item $J_i\cap J_k = \emptyset$ for all $j,k\in [L]$,
    \item $\sum_{k=1}^L |J_k| = d$,
    \end{enumerate}
    for which if we define $x^{(k)}$ 
    via 
    \begin{equation}
    x^{(k)}_i = \begin{cases} x_i\text{ if } i\in J_k, \\
    0 \text{ else},
    \end{cases}
    \end{equation}
    then $\mathbf{B}^{(k)}(x) = \mathbf{B}^{(k)}\!\left(x^{(k)}\right)$---i.e., all 
    functions in $\mathbf{B}^{(k)}$ depend only on elements of $x$ with indices in the set $J_k$.
\end{enumerate}
With these assumptions, one can see that there exists a TN representation
\begin{align}\label{eq:TN-rep}
    f(x) &= \sum_{i=1}^Dc_ib_i(x)=\mathbf{C}\cdot\bigotimes_{k=1}^L\mathbf{B}^{(k)}(x) \nonumber\\
    &=\vcenter{\hbox{\includegraphics[height=0.25\linewidth]{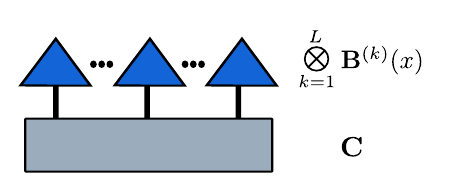}}}.
\end{align} Note that the number of function dictionaries $L$ can be logarithmic 
in the number of functions $D$.\\

\noindent \textbf{Example:} Consider the bivariate function  $f(x_1,x_2) = 1 + 2\cos(x_1) -2\cos(x_1+x_2)$. Define $\mathbf{B}^{(1)}(x) = [1~\cos(x_{1})~\sin(x_{1})]^\top$ and $\mathbf{B}^{(2)}(x) = [1~\cos(x_{2})~\sin(x_{2})]^\top$. Using the fact 
that $\cos(x_1+x_2) = \cos(x_1)\cos(x_2)-\sin(x_1)\sin(x_2)$ allows us to show that, 
indeed,
\begin{align}
f(x_1,x_2) &= 1 + 2\cos(x_1) \jetext{- 2\cos(x_1)\cos(x_2)} \nonumber\\ &\, \jetext{+2\sin(x_1)\sin(x_2)} \nonumber\\ 
&= B^{(1)}_0(x)B^{(2)}_0(x) + 2B^{(1)}_1(x)B^{(2)}_0(x) \nonumber\\&\,\jetext{- 2B^{(1)}_1(x)B^{(2)}_1(x) +2B^{(1)}_2(x)B^{(2)}_2(x)} \nonumber\\
&= \sum_{i_1,i_2 = 0}^2\mathbf{C}_{i_1, i_2}B^{(1)}_{i_1}(x)B^{(2)}_{i_2}(x) 
\nonumber\\
&= \mathbf{C}\cdot\bigotimes_{k=1}^2\mathbf{B}^{(k)}(x),
\end{align}
with 
\begin{equation}
    \mathbf{C}_{0,0} = 1,~ \jetext{\mathbf{C}_{1,0}} = 2, ~\mathbf{C}_{1,1}= \jetext{-2},~\mathbf{C}_{2,2} =\jetext{2} \jetext{.}
\end{equation}

When a function admits such a TN representation, \jetext{learning the function translates} to learning the corresponding coefficient tensor $\mathbf{C}$ for given local function dictionaries. With increasing dimension of the input, \jetext{the number of basis functions and, therefore,} \jetext{the coefficient entries to be learned increase} exponentially, suffering from `curse of dimensionality'. For this reason, one might only hope for efficient learning when the coefficient tensor is sparse or sufficiently concentrated. However, given that \jetext{the function is fully specified by the coefficient tensor, a natural idea} is to try and exploit the TCI algorithm to learn it. Indeed, \jetext{this should work} efficiently whenever $\mathbf{C}$ is approximable by $O(\text{poly}(L))$ bond dimensions (keeping in mind that $L\leq d$).

The immediate problem is that we only have the ability to query $f$, and \textit{not} the ability to query $\bf{C}$, which is required by TCI. As such, we \textit{cannot} simply apply TCI on the coefficient tensor. \jetext{In the following, however,} we show a simple trick which circumvents this obstacle, and allows us to exploit TCI in this context.

\subsubsection{A key trick: From function values to tensor entries}

Recall that under our assumptions the target function $f$ 
has an efficient  tensor network representation as in Eq.~\eqref{eq:TN-rep}, and is fully 
specified by the coefficient tensor $\mathbf{C}$, to which unfortunately we do not 
have direct access. In this section we show a simple trick, which allows us to define a new ``virtual tensor'' $\mathbf{F}$, for which the following holds true.
\begin{enumerate}
\item We \textit{can} access every entry of $\mathbf{F}$ using query access to $f(x)$.
\item We can (approximately) recover $\mathbf{C}$ from $\mathbf{F}$.
\end{enumerate}
In particular, we do the following: \\

\noindent\textbf{(a) Choose a product grid of query points.}
For each input block $J_k$, choose $M^{(k)}$ auxiliary inputs
\begin{equation}
    A^{(k)}=\left\{a^{(k)}_{j_k}\in\mathbb{R}^{|J_k|}\,\middle|\,j_k\in[M^{(k)}]\right\}.
\end{equation}
For a multi-index $\mathbf{j}=(j_1,\ldots,j_L)$, let $a_{\mathbf{j}}\in\mathbb{R}^d$ be the global input obtained by placing $a^{(k)}_{j_k}$ in the coordinates indexed by $J_k$. Since $\{J_1,\ldots,J_L\}$ is a partition of $[d]$, every choice of $\mathbf{j}$ specifies one unambiguous global $d$-dimensional input. Thus, the local sets define a product grid of $M=\prod_{k=1}^L M^{(k)}$ query points.\\

\noindent\textbf{(b) From the grid to a tensor.} First, we collect 
auxiliary input vectors indexed by multi-index $\mathbf{j}$,
\begin{equation}
\{a_{j_1,\ldots,j_L} = [a_{j_1}^{(1)},\ldots,a_{j_L}^{(L)}] \in \mathbb{R}^d \,|\, j_i \in [M^{(i)}] \}_{\mathbf{j}}.
\end{equation}
Now, given these auxiliary input vectors, we evaluate each $k$'th function dictionary $\mathbf{B}^{(k)}$ on \jetext{the points $a^{(k)}_{j_k}$}, and collect the results into the matrix $V^{(k)}$, in other words, we construct
\begin{align}\label{eq:basis_matrices}
    V^{(k)}(\mathbf{B}^{(k)},a^{(k)}) = \big[
        \mathbf{B}^{(k)}(a^{(k)}_1) ~ \mathbf{B}^{(k)}(a^{(k)}_2) ~\cdots ~ \mathbf{B}^{(k)}(a^{(k)}_{M^{(k)}})\big].
\end{align} 
where the size is $Q^{(k)}\times M^{(k)}$. The column vector $\mathbf{B}^{(k)}(a^{(k)}_{j_k})$ is obtained by evaluating the functions in $\mathbf{B}^{(k)}$ on input $a^{(k)}_{j_k}$. Equivalently, $V^{(k)}(\mathbf{B}^{(k)},a^{(k)})_{i_kj_k}=B^{(k)}_{i_k}(a^{(k)}_{j_k})$. 
Finally, with this in hand, we define the virtual tensor $\mathbf{F}$ by contracting the coefficient tensor $\mathbf{C}$ with these $L$ matrices, i.e., 
\begin{align}
\mathbf{F} &:= \mathbf{C} \cdot \bigotimes_{k=1}^LV^{(k)}(\mathbf{B}^{(k)},a^{(k)}), \nonumber\\
&=\vcenter{\hbox{\includegraphics[height=0.25\linewidth]{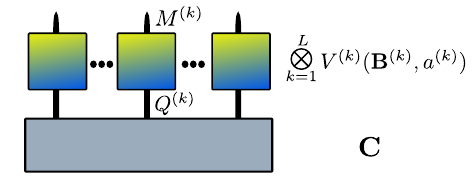}}}
\end{align}

Now, we can make the following observations.
\begin{observation}[Function values as tensor entries] The entries of virtual tensor $\mathbf{F}$ are equal to the values of the function $f$ on vectors $\{a_{j_1,\ldots,j_L}\}$, i.e.,
\begin{equation}\mathbf{F}_{j_1, \ldots , j_L} = f(a_{j_1,\ldots,j_L}).\end{equation}
\end{observation}
As such, unlike the coefficient tensor $\mathbf{C}$, we \textit{do} have access to the entries of $\mathbf{F}$ via query access to $f(x)$, and can therefore run TCI to obtain a low bond-dimension approximation $\tilde{\mathbf{F}}$ to $\mathbf{F}$. Of course, ultimately we are not interested in $\mathbf{F}$, and would like an approximation to $\mathbf{C}$! To this end, for all matrices $V^{(k)}$ let us define the pseudoinverse $(V^{(k)})^+$, and the tensor $\tilde{\mathbf{C}}$ as the one obtained by contracting $\tilde{\mathbf{F}}$ with the $L$ pseudoinverses $(V^{(k)})^+$, i.e., 
\je{we have}
\begin{align}
\tilde{\mathbf{C}} &= \tilde{\mathbf{F}}\cdot \bigotimes_k(V^{(k)})^{+} \nonumber\\
&= \vcenter{\hbox{\includegraphics[height=0.25\linewidth]{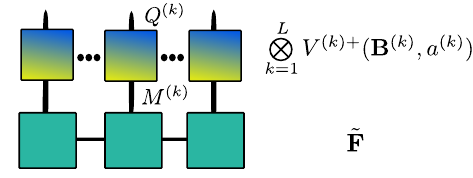}}}
\end{align}
We can then make the following observation:

\begin{observation}[Exact recovery from full-row-rank designs] When $Q^{(k)} < M^{(k)}$, and $V^{(k)}$ has a full-row rank, then $V^{(k)}V^{(k)+} = I$. Therefore, if $\tilde{\mathbf{F}} =\mathbf{F}$, then $\tilde{\mathbf{C}} = \mathbf{C}$. 
\end{observation}
Of course, in practice we do not have $\tilde{\mathbf{F}} =\mathbf{F}$ or exact inverses, and so $\tilde{\mathbf{C}}$ will only be an approximation to $\mathbf{C}$. We will make these approximation errors precise shortly, however before that we first formalize the algorithm that we have now implicitly developed.

\subsubsection{FTCI algorithm}

\begin{figure*}
    \centering
    \includegraphics[width=1.0\linewidth]{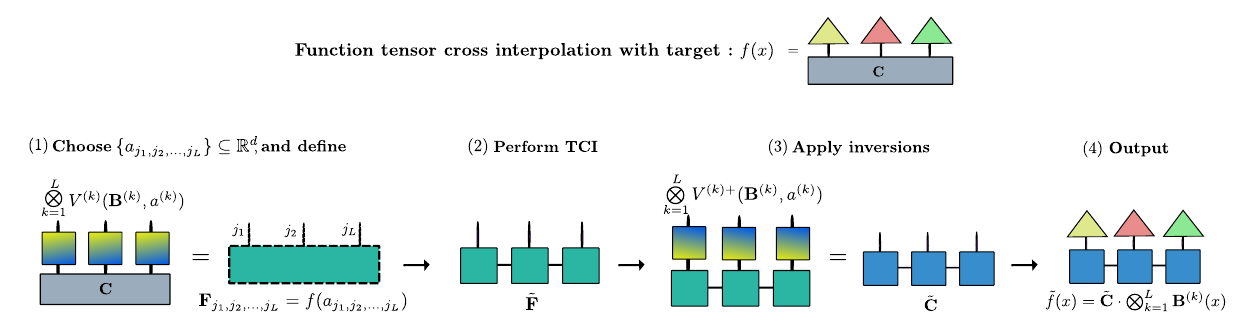}
    \caption{Function tensor cross interpolation. We are given basis function vectors (colored triangles), and oracle access to target function $f$.  (1) Construct a ``virtual" tensor $\mathbf{F}$ that can be accessed by querying auxiliary input vectors $\{a_{j_1 ,j_2,\ldots, j_L}\}$ on $f$. Note that contracting \jetext{the matrices $V^{(k)}$} to coefficient tensor $\mathbf{C}$ does not occur in practice. (2) Run the TCI algorithm to this tensor to obtain an approximate MPS tensor $\tilde{\mathbf{F}}$. (3) Calculate the pseudoinverses $V^{(k)+}$s using basis 
    functions and \jetext{the} auxiliary input set. Contract them \jetext{with $\tilde{\mathbf{F}}$ to obtain $\tilde{\mathbf{C}}$,} an approximate coefficient tensor. (4) Using the given basis function vectors and approximated coefficient, we obtain an approximated function in the TN form.}
    \label{fig:FTCI}
\end{figure*}

Given the definitions, notation and observations of the previous section, a high-level description of our proposed FTCI algorithm is as follows:

\begin{enumerate}
\item Choose a set of \jetext{auxiliary} input vectors  $a^{(k)}=\{a^{(k)}_{j_k}\in \mathbb{R}^{|J_k|}\,|\, j_k \in [M^{(k)}]\}$.
\item Perform TCI on the implicitly defined virtual tensor $\mathbf{F}$ and obtain a low bond-dimension MPS approximation $\tilde{\mathbf{F}}$.
\item Contract $\tilde{\mathbf{F}}$ with the pseudoinverse matrices $(V^{(k)})^+$ to obtain $\tilde{\mathbf{C}}$, a low bond-dimension MPS approximation to the coefficient tensor $\mathbf{C}$ \jetext{specifying} $f(x)$.
\item Output the function 
\begin{equation}
    \tilde{f}(x) = \tilde{\mathbf{C}}\cdot \bigotimes_{k=1}^L\mathbf{B}^{(k)}(x).
\end{equation}
\end{enumerate}
A more precise version of this algorithm is given as Algorithm~\ref{alg:FTCI}, which is illustrated in Fig.~\ref{fig:FTCI}.

\subsubsection{Computational and sample complexity}

\jetext{At a high level}, 
FTCI consists of 
two 
steps: (i) TCI on the virtual tensor $\mathbf{F}$, and (ii) Applying local inversions to $\tilde{\mathbf{F}}$. \jetext{Only the first step requires samples from $f$; therefore,} the sample complexity for FTCI is the same as TCI. More specifically, the sample complexity for TCI scales as $O(\chi^2M_{\rm max}^2L)$~\footnote{The power of $M_{\rm max}$ could be different by detailed algorithm used during TCI.}, where $L$ is the number of function dictionaries, $\chi$ is the maximum bond dimension \jetext{for the TCI algorithm}, and $M_{\mathrm{max}}$ is the maximum local dimension of virtual tensor $\mathbf{F}$, which corresponds to the maximum number of auxiliary points per function dictionaries. 

The algebraic cost for step (i) scales with $O(\chi^3M_{\rm max}^2L)$, and the inversion step (ii) introduces an additional algebraic cost. However, thanks to the tensor product structure, applying the inversion process takes only $O(\chi^2Q_{\rm max}M_{\rm max}
L)$ algebraic operations. Also, obtaining pseudo-inverses of \jetext{the matrices $V^{(k)}$} takes $ O(M_{\rm max}^3L)$ operations. 

In conclusion, FTCI does not increase the sample complexity from TCI subroutine, and the additional algebraic cost also scales polynomially with respect to $L$ (and therefore $d$), as long as \jetext{the values $M^{(k)}$} do not scale exponentially. Therefore, FTCI works 
efficiently whenever $\chi = O(\text{poly}(L))$, just like TCI. 

\begin{figure*}[t]
\centering
\begin{algorithm}[H]
\caption{Function tensor cross interpolation (FTCI)}
\label{alg:FTCI}
\DontPrintSemicolon

\KwIn{
Oracle access to $f:\mathbb{R}^d\to\mathbb{R}$\\
Partition $\{J_k\}_{k=1}^L$ of $[d]$\\
Local basis vectors $\left\{\mathbf{B}^{(k)}:\mathbb{R}^{|J_k|}\to\mathbb{R}^{Q^{(k)}}\right\}_{k=1}^L$\\
Product grid of auxiliary points $A^{(k)}=\left\{a^{(k)}_{j_k}\in\mathbb{R}^{|J_k|}\,\middle|\,j_k\in[M^{(k)}]\right\}_{k=1}^L$\\

Max bond dimension $\chi_{\max}$ and TCI tolerance $\tau$.
}
\KwOut{Coefficient tensor $\tilde{\mathbf{C}}$ in MPS form and predictor $\tilde f(x)$.}

\BlankLine

\BlankLine
\tcp{Step 1: define the implicit $L$-leg tensor $\mathbf{F}$ accessible by queries to $f$}
\SetKwFunction{Oracle}{OracleF}
\SetKwProg{Fn}{Function}{:}{}
\Fn{\Oracle{$j_1,\ldots,j_L$}}{
  Assemble $a_{j_1, \ldots , j_L}\leftarrow [a_{j_1}^{(1)},\ldots,a_{j_L}^{(L)}]\in\mathbb{R}^d$\;
  \Return $f\!\left(a_{j_1, \ldots , j_L}\right)$\;
}

\BlankLine
\tcp{Step 2: run TCI on $\mathbf{F}$ without explicitly forming it}
$\tilde{\mathbf{F}} \leftarrow \textsc{TCI}\Big(\Oracle;\ (M^{(1)},\ldots,M^{(L)}),\ \tau,\ \chi_{\max}\Big)$\;

\BlankLine
\tcp{Step 3: recover coefficients by applying local pseudo-inverses on each leg}
\For{$k\leftarrow 1$ \KwTo $L$}{
  Initialize $V^{(k)}\in\mathbb{R}^{Q^{(k)}\times M^{(k)}}$~(Eq.\ref{eq:basis_matrices})\;
  \For{$j_k\leftarrow 1$ \KwTo $M^{(k)}$}{
    Set the $j_k$-th column: $V^{(k)}_{:,j_k} \leftarrow \mathbf{B}^{(k)}(a_{j_k}^{(k)})$\;
  }
  Compute pseudo inverse $(V^{(k)})^{+}$ 
}
$\tilde{\mathbf{C}} \leftarrow \tilde{\mathbf{F}} \cdot \bigotimes_{k=1}^{L}(V^{(k)})^{+}$\;

\BlankLine
\tcp{Step 4: define the learned function}
\SetKwFunction{Predict}{Predict}
\Fn{\Predict{$x\in\mathbb{R}^d$}}{
  Parse $x=[x^{(1)},\ldots,x^{(L)}]$ with $x^{(k)}=[x_i]_{i\in J_k}$\;
  \Return $\tilde{\mathbf{C}}\cdot \bigotimes_{k=1}^{L}\mathbf{B}^{(k)}(x^{(k)})$\;
}

\Return $(\tilde{\mathbf{C}},\ \tilde f\equiv \Predict)$\;
\end{algorithm}
\end{figure*}

\subsubsection{Error propagation}

It is important to stress that the method proposed is not only expected to be
heuristically very practical. But indeed, the predictions can be equipped with
precise guarantees.
Here, we discuss the error propagation of the method.
Note that
\begin{align}\label{eq:error_propagation}
    \|\mathbf{C}-\mathbf{\tilde{C}}\|_2 &=   \left\|\mathbf{F}\cdot\bigotimes_{k=1}^L{V}^{(k)+} - \mathbf{\tilde{F}}\cdot\bigotimes_{k=1}^L{V}^{(k)+}\right\|_2 \nonumber\\&\leq \|\mathbf{F} - \mathbf{\tilde{F}}\|_2\norm{\bigotimes_{k=1}^L{V}^{(k)+}},
\end{align}
where $\norm{\cdot}$ stands for spectral norm. This means that the error of the approximated coefficient tensor could blow up with an unfortunate choice of the auxiliary vectors $a_{j_1\ldots j_L}$. Therefore, one needs to choose auxiliary points that make the condition numbers of \jetext{the matrices $V^{(k)}$} small, or equivalently, their smallest singular values large. This problem could be overcome by oversampling, and this is why we make $M^{(k)}$ larger than $Q^{(k)}$. In fact, for trigonometric basis functions that are relevant for the case of quantum functions, we have a concentration result for the spectral norm, which we will present in detail in Section~\ref{ss:Error_analysis_for_quantum}\jetext{.}

\subsubsection{Short comparison between FTCI and quantics TCI.}

We note here that within the TCI community an algorithm named \textit{quantics TCI} (QTCI) has already been developed for function approximation~\cite{ritter2024quantics}. Specifically, given a rescaled function $f :[0,1)^d \rightarrow \mathbb{R}$, one generates the \textit{quantics} representation $\mathbf{F}$ of $f$ by quantizing the input domain with \edit{$d \times R$} bits, and assigning
\begin{eqnarray}
    &&\mathbf{F}(i_{1,1},\ldots ,i_{1,R},i_{2,1},\ldots ,i_{2,R}\ldots, i_{d,1}, i_{d,R}) \\
    &&\mapsfrom
    f\left(\sum_{b=1}^R\frac{i_{1,b}}{2^b},\ldots,\sum_{b=1}^R\frac{i_{d,b}}{2^b}\right).
    \nonumber
\end{eqnarray}
Here, $R$ can be thought of as the resolution of the discretized domains. Now, one can use TCI algorithm to approximate $\mathbf{F}$, and if the $\mathbf{F}$ indeed has small bond dimensions, then one can learn this function efficiently with exponentially fine-grained resolution. 

One trivial difference between FTCI and QTCI is that QTCI only recovers finite points of the function, while FTCI recovers a continuous domain function. However, this does not immediately mean that FTCI is more expressive than QTCI by virtue of its continuous domain support, because by growing $R$ in QTCI, one can easily obtain a grid that is fine enough to cover the whole domain within machine precision. 

Instead, one can think of the quantics representation of $f$ as the TN representation where basis function dictionaries are 
\begin{equation}
\begin{pmatrix}
1-b^{(k)}_j(x)\\
b^{(k)}_j(x)  
\end{pmatrix}, 
\end{equation}
where $b^{(k)}_{j}: [0,1)^d\rightarrow\{0,1\}$ is the function that outputs $j$'th bit of $k$'th element of $x$. And QTCI learns this representation. Meanwhile, FTCI's objective is to learn a coefficient \textit{given} some function dictionaries that are not constrained to binary basis functions.

\section{Learning quantum functions}\label{sec:learning-q-functions}

The quantum function learning algorithm directly follows from the previous two sections. We apply FTCI to quantum functions in the form given in Eq.~\eqref{eq:TN_quantum_function}, using basis function vectors shown in Eq.~\eqref{eq:canonical_basis}. However, some quantum functions can have pre-processing functions that make different basis function vectors depend on the same element of the input at the same time. This violates Assumption 4 of Section~\ref{ss:assumptions}, which states that all basis vectors should be independent of each other with respect to a disjoint set of inputs. Luckily, however, we have direct access to the basis part for quantum functions by changing the pre-processing functions that determine the rotational input to encoding gates. 

More specifically, the trick is to construct a new virtual tensor by just changing the original pre-processing function set $\mathcal{P}$ to $\mathcal{P'} :=\{ \phi^{(j)}: a \mapsto a_j~ |~j\in [|\mathcal{G}|]\}$, with auxiliary inputs $a \in \mathbb{R}^{|\mathcal{G}|}$. Using this modification, the quantum function satisfies all the assumptions needed for FTCI. Now our virtual tensor is
\begin{equation}
    \mathbf{F'_Q} = \mathbf{C}\cdot\bigotimes_{k=1}^LV'^{(k)},
\end{equation}where $L:=|\mathcal{G}|$, and
\begin{align}\label{eq:quantum_design}
     &V'^{(k)} :=\sqrt{\frac{1}{D^{(k)}+1}}\left[\mathbf{\Phi}'^{(k)}(a^{(k)}_{1}),\cdots,\mathbf{\Phi}'^{(k)}(a^{(k)}_{M^{(k)}})  \right]
     \end{align}
     
     \begin{align}
    &\mathbf{\Phi}'^{(k)}(a^{(k)}_{j_k}) := \sqrt{\frac{1}{D^{(k)}+1}}\begin{pmatrix}
        1 \\ \cos\omega^{(k)}_1 a_{j_k}^{(k)} \\ \sin \omega^{(k)}_1 a_{j_k}^{(k)} \\ \vdots \\ \cos\omega^{(k)}_{D^{(k)}}a_{j_k}^{(k)} \\ \sin \omega^{(k)}_{D^{(k)}} a_{j_k}^{(k)}
    \end{pmatrix} \nonumber\\
    &a_{j_1, \ldots , j_L} = [a^{(1)}_{j_1},~\ldots~,a^{(L)}_{j_L}],\quad a^{(k)}_{j_k} \in [0,P^{(k)}],~j_k\in[M^{(k)}]
\end{align}
with frequency sets that are given by Eq.~\eqref{eq:frequency_set}.

\begin{figure*}
    \centering
    \includegraphics[width=1.0\linewidth]{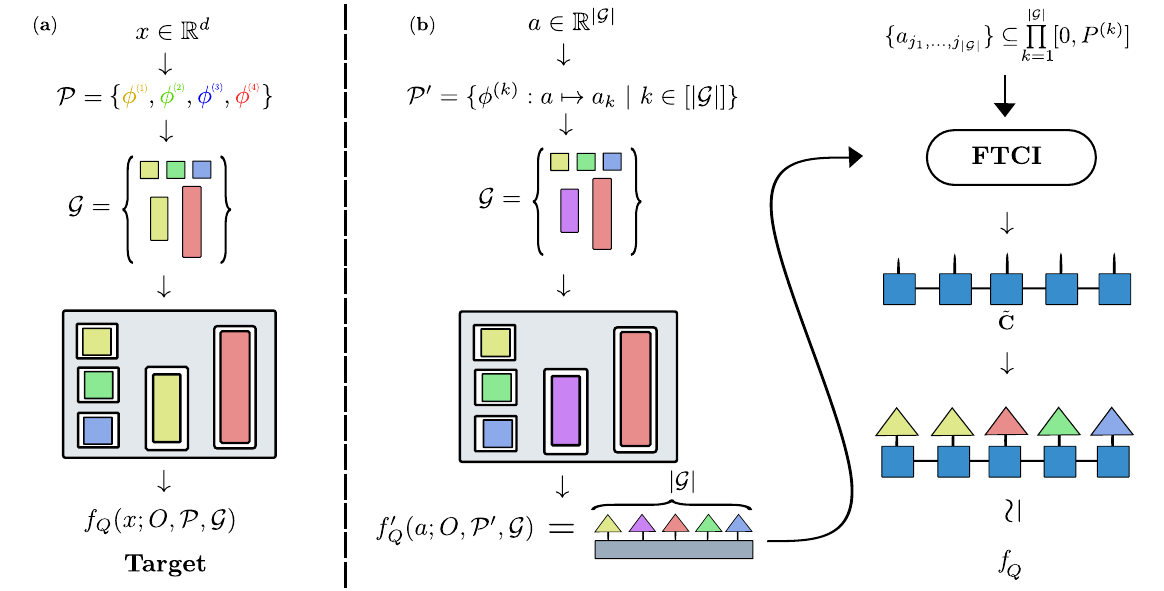}
    \caption{Quantum function learning with FTCI. (a) We are given a quantum function with an input-dependent unitary generator set $\mathcal{G}$ and a set of pre-processing functions $\mathcal{P}$. For illustration we depict $|\mathcal{G}|= 5$, and $|\mathcal{P}|= 4$ case. (b) To apply FTCI, we replace $\mathcal{P}$ by a modified pre-processing set $\mathcal{P}'$. We use this modified quantum function $f'_Q$ as an input to FTCI algorithm with auxiliary inputs whose components are sampled from intervals $[0,P^{(k)}]$. We get the approximated coefficient tensor and use the original function dictionaries using $\mathcal{P}$, and $\mathcal{G}$. Choosing the domain ranges $P^{(k)}$ as specified in the main text ensures that the inversion-induced error in FTCI does not blow up for a modest number of auxiliary samples. This figure illustrates the process of learning the canonical form Eq.~\eqref{eq:TN_quantum_function}.} 
    \label{fig:QFTCI}
\end{figure*}

In particular, the elements of the tensor $\mathbf{F'_Q}_{j_1, \ldots , j_L} = f'_Q(a_{j_1, \ldots , j_L})$, are accessible by querying the quantum function $f'_Q$, where 
\begin{equation}
    f'_Q(a; O,\mathcal{P',G}): [0,P^{(1)}]\times\cdots\times[0,P^{(L)}] \rightarrow \mathbb{R}.
\end{equation} 
Note that the only modifications required to obtain $\mathbf{F'_Q}$ were (a) changing the pre-processing functions and (b) using $L$-dimensional auxiliary inputs instead of $d$-dimensional ones. As such, the coefficient tensors of $f'_Q$ and $f_Q$ are the same. Therefore, after recovering the approximate coefficient tensor $\tilde{\mathbf{C}}$ by running FTCI on $f'_Q$, we use the original basis function vectors \jetext{$\Phi^{(k)}(x)$} to recover $\tilde{f}_Q$. See Fig.~\ref{fig:QFTCI}, and Algorithm~\ref{alg:quantum_function_ftci} for the pseudocode.

The above algorithm learns a canonical TN representation where the length 
of the tensor $L = |\mathcal{G}|$, which is the number of input-dependent unitaries we use. In many practical 
settings however, like data-reuploading models in QML, or Hamiltonian simulation, one uses the same 
pre-processing function for many different encoding unitaries. In that case, as we discussed in Sec.~\ref{ss:TN_repre_of_quantum}, we can use a shorter representation than the canonical representation by merging basis function vectors sharing the same pre-processing functions. In this case, we set 
\begin{equation}
\mathcal{P'} :=\{ \phi^{(k)} : a \mapsto a_k~ |~k\in [|\mathcal{P}|]\}, 
\end{equation} 
and input-dependent unitaries that share the same \jetext{pre-processing} function $\phi^{(k)}$ \jetext{take} the same auxiliary input $a^{(k)}$ when we query the quantum functions. In this way we obtain the tensor of length $L= |\mathcal{P}|$, with merged function dictionaries.

\begin{figure*}[t]
\centering
\begin{algorithm}[H]
\caption{Learning a quantum function via FTCI }
\label{alg:quantum_function_ftci}
\DontPrintSemicolon
\KwIn{
Quantum-function oracle access $f_Q(x;O,\mathcal{P},\mathcal{G})$\\
Canonical function dictionaries $\Phi^{(k)}$ (Eq.~\eqref{eq:canonical_basis}).\\
Frequency sets $\Omega^{(k)}$s (Eq.~\eqref{eq:frequency_set})\\
  $\{M^{(k)}\}_{k=1}^L,\qquad M^{(k)}\ge 2D^{(k)}+1$ \\
Max bond dimension $\chi_{\max}$ and TCI tolerance $\tau$.
}
\KwOut{Approx.\ coefficient tensor $\tilde{\mathbf{C}}$ in MPS form and learned $\tilde f_Q(x)$.}
\BlankLine

\tcp{Step 0: define the auxiliary-quantum function $f'_Q$}
\SetKwFunction{QOracle}{QOracle}
\SetKwProg{Fn}{Function}{:}{}
\Fn{$f'_Q(a_1,\ldots,a_L)$}{
  \tcp{Set pre-processing maps $\phi'_k:a\mapsto a_k$ and evaluate the same circuit}
  \Return $f_Q\left(x=a;~O,~\mathcal{P}'=\{\phi'_k\}_{k=1}^L,~\mathcal{G}\right)$
}
  
\BlankLine
\tcp{Step 1: construct auxiliary basis vectors $v^{(k)}$ (Trigonometric-basis)}
\For{$k\leftarrow 1$ \KwTo $L$}{
  Set 
  \begin{align}
      \mathbf{B}^{(k)}(x) &\leftarrow v^{(k)}(x) = \sqrt{\frac{1}{D^{(k)}+1}}
      \begin{pmatrix}
        1
        ~\cos(\omega^{(k)}_1 x)
        ~\sin(\omega^{(k)}_1 x)
        ~\cdots~
        ~\cos(\omega^{(k)}_{D^{(k)}} x)
        ~\sin(\omega^{(k)}_{D^{(k)}} x)
      \end{pmatrix}^{\top}
    \end{align}
}

\BlankLine
\tcp{Step 2: Sample auxiliary input set $\{\{a^{(k)}_{j_k}\}~|~ k\in[L]\}$}
\SetKwFunction{AuxSet}{AuxSet}
\Fn{\AuxSet{$\{~\{\omega^{(k)}_j\}~|~k\in[L]~\}$}}{
  \For{$k\leftarrow 1$ \KwTo $L$}{
    $\gamma^{(k)} \leftarrow \min\Big\{\,\min_j|\omega^{(k)}_j|,\ \min_{j\neq h}|\omega^{(k)}_j-\omega^{(k)}_h|~\Big\}$\\
    Choose $P^{(k)} > 2\pi/\gamma^{(k)}$\\
    \For{$j_k\leftarrow 1$ \KwTo $M^{(k)}$}{
    Sample $a^{(k)}_{j_k}\sim\mathrm{Unif}([0,P^{(k)}])$.}
  }
  \Return $\{\{a^{(k)}_{j_k}\}~|~k\in[L]\}$
  }
\BlankLine
\tcp{Step 3: Run FTCI}
$\tilde{\mathbf{C}}\leftarrow \text{FTCI}(f'_Q,~\{\{k\}~|~k\in[L]\},~ \{v^{(k)}\},~\{~\{a^{(k)}_{j_k}\}~|~k\in[L]\},~\chi_{\rm max},~\tau )$ 

\BlankLine
\tcp{Step 4: output predictor for the original quantum function $f_Q(x)$}
\SetKwFunction{Predict}{Predict}
\Fn{\Predict{$x\in\mathbb{R}^d$}}{
  \tcp{Evaluate the original function dictionaries $\Phi^{(k)}(x)$ from Eq.~\eqref{eq:canonical_basis}}
  \Return $\tilde{\mathbf{C}}\cdot \bigotimes_{k=1}^{L}\Phi^{(k)}(x)$\;
}
\Return $ \tilde f_Q\equiv \Predict$\;
\end{algorithm}
\end{figure*}

\subsection{Error analysis}\label{ss:Error_analysis_for_quantum}

Once again, we provide an error analysis. For quantum function learning, there are three sources of error. The first is the statistical noise from using a quantum system. When we query one element $\mathbf{F'_Q}(i_1, \ldots , i_L) $, we get $\mathbf{F'_Q}(i_1, \ldots,  i_L) + \varepsilon$ for each query in reality. The second comes from the TCI error, which is a fundamental error from low-rank approximability of the target function, and suboptimality of TCI itself. The final one is 
the error from the inversion process. 

For the first and second error sources, they are analyzed in the setting of \emph{noisy TCI}. Noisy TCI addresses the case where one seeks to approximate a tensor $\mathbf{T}$ but has access only to a noisy version $\mathbf{T} + \bm{\mathcal{E}}$. In the quantum setting, one may interpret $\bm{\mathcal{E}}$ as a noise tensor induced by shot noise, whose entries are on the order of $O(1/\sqrt{S})$, where $S$ is the number of samples to estimate the expectation value.

Importantly, in this work we do not separately analyze shot-noise error beyond this reduction. We treat it as part of the noisy-oracle model required for noisy TCI, and directly rely on existing guarantees in the noisy TCI literature~\cite{qin2022error}, and efforts to improve it~\cite{sakaue2025adaptive}. However, we expect moderate error scaling when we choose a proper truncation parameter for the pseudo-inverse in the TCI algorithm. See Appendix~\ref{app:Noisy} for discussion.

The final error, originating from the inversion step, is a new source of error introduced by FTCI. It is governed by the spectral norm of the inversion matrices, as in Eq.~\eqref{eq:error_propagation}. In the quantum setting, the inversion matrices $V^{(k)+}$ are determined by trigonometric functions, as in Eq.~\eqref{eq:quantum_design}, and we provide a concentration bound for this special case.

\begin{theorem}[Inversion-error bound for trigonometric dictionaries]\label{thm:error_supression}
Let the target function tensor be $\mathbf{F} = \mathbf{C}\cdot \bigotimes_{k=1}^L V'^{(k)}(a^{(k)})$, where $V'^{(k)}(a^{(k)})$ is given in Eq.~\eqref{eq:quantum_design}. Let the frequency set be $\{\omega^{(k)}_i\}_{i=1}^{D^{(k)}}$, and let the auxiliary points $a^{(k)}_1,\ldots,a^{(k)}_{M^{(k)}} \stackrel{\text{i.i.d.}}{\sim}\mathrm{Unif}[0,P^{(k)}]$. Assume that the domain ranges satisfy $P^{(k)} > 2\pi/\gamma^{(k)}$, where
\begin{equation}
0< \gamma^{(k)} \leq \min\Big\{\,\min_j|\omega^{(k)}_j|,\ \min_{i\neq j}|\omega^{(k)}_i-\omega^{(k)}_j|\,\Big\}.
\end{equation}
Let $\tilde{\mathbf{C}} = \tilde{\mathbf{F}}\cdot\bigotimes_{k=1}^L\bigl( V'^{(k)}(a^{(k)})\bigr)^+$; then, with probability at least $1-\delta$,
\begin{align}\label{eq:error_bound}
    \|\mathbf{C} - \tilde{\mathbf{C}}\|_2  \leq \left(\frac{C_{\max}}{(1-\varepsilon)}\frac{D_{\max}+1}{M_{\min}}\right)^{L/2}\norm{\mathbf{F}-\tilde{\mathbf{F}}}_2,
\end{align}
where
\begin{equation}\label{eq:auxiliary_number}
    M_{\min} \geq \frac{2C_{\max}(D_{\max}+1)}{\varepsilon^2}\left(\log(2D_{\max}+1) + \log\left(\frac{L}{\delta}\right) \right).
\end{equation}
Here, $M_{\min} = \min_k M^{(k)}$, $D_{\max} = \max_k D^{(k)}$, the constant $C_{\max} := \max_k C(P^{(k)},\gamma^{(k)})$ with $C(P,\gamma) = 2P/(P-2\pi/\gamma)$ as defined in Appendix~\ref{app:thm1_proof}, $\delta \in (0,1)$, and $\varepsilon \in (0, \sqrt{3}-1]$.
\end{theorem}

We defer the proof to Appendix~\ref{app:thm1_proof}. In other words, the more auxiliary points we use, the better the accuracy we obtain. Moreover, the required number of auxiliary points does not increase exponentially in $L$, and therefore preserves the efficiency of TCI. Note that if we choose $M_{\min}$ as in Eq.~\eqref{eq:auxiliary_number} and substitute it into the right-hand side of Eq.~\eqref{eq:error_bound}, then
\begin{equation}
    \frac{\varepsilon^2}{2B(1-\varepsilon)} \leq 1,\quad \text{for}~\varepsilon\in(0,\sqrt{3}-1],
\end{equation}
with $B = \log(2D_{\max}+1) + \log(L) + \log(1/\delta) > 1$. Therefore, $M_{\min} = O({D_{\max}}\log(L/\delta))$ suffices to suppress the inversion-induced error for large $L$, and bounds the amplification of measurement noise by the inversion step. In practice, setting $M^{(k)} = 2(2D^{(k)}+1)$---which corresponds to doubling the local dimension---was sufficient to achieve good performance in all of our noiseless numerical experiments.

\section{Conditions for learning quantum functions using FTCI}\label{sec:conditions_for}

In this section, we ask when learning quantum functions can be efficient with our approach. To address this, we first identify conditions under which FTCI itself can run efficiently and yield accurate reconstructions. We then return to the quantum-function setting and discuss when these conditions can be expected to hold for coefficient tensors, \jetext{and connect them to quantum process properties} for the special case.

\subsection{General conditions for FTCI efficiency}\label{sec:ftci_efficiency}

We begin with the tensor properties that govern the accuracy and cost of the interpolation step.
Tensor cross interpolation is essentially a compression algorithm: given a target tensor, it constructs an MPS approximation with maximum bond dimension $\chi$. Consequently, the best achievable approximation error for a fixed $\chi$ is ultimately governed by the compressibility of the target tensor---equivalently, by its ``true'' MPS bond dimension $\chi_{\mathrm{target}}$. This optimal error (for a given $\chi$) is attained by SVD-based truncation, whereas TCI provides a computationally efficient, though generally suboptimal, alternative.

Function TCI uses TCI as a subroutine and therefore inherits these traits. In FTCI, the target object is the (implicit) function tensor $\mathbf{F}$. Importantly, $\mathbf{F}$ is constructed from $\mathbf{C}$ via local operations. Therefore,
\begin{equation}
    \chi(\mathbf{F}) \le \chi(\mathbf{C}),
\end{equation}
where $\chi(\cdot)$ denotes the maximum bond dimension of the tensor when represented in MPS form. Consequently, FTCI is efficient when the \textit{coefficient} tensor can be approximated well using $\chi = O(\mathrm{poly}(L))$, where $L$ is the number of local basis vectors.

More formally, let us consider normalized coefficient tensors with $\norm{\mathbf{C}}_2=1$. Let $S^{(\alpha)}(\mathbf{C})$ denote the R\'enyi-$\alpha$ entropy of the squared entries of $\mathbf{C}$ (viewed as a vector), i.e.,
\begin{equation}
    S^{(\alpha)}(\mathbf{C}) := \frac{1}{1-\alpha}\log\sum_{i_1,\ldots,i_L}|\mathbf{C}_{i_1,\ldots,i_L}|^{2\alpha}.
\end{equation}
We also define the entanglement of $\mathbf{C}$ with respect to a bipartition $(A:B)$ as
\begin{equation}
    E^{(\alpha)}_{(A:B)}(\mathbf{C}) := S^{(\alpha)}\!\bigl(\vec{\sigma^2}(\mathbf{C}_{(A:B)})\bigr),
\end{equation}
where $\mathbf{C}_{(A:B)}$ is the matrix obtained by reshaping $\mathbf{C}$ according to the index partition $(A:B)$, and $\vec{\sigma^2}(A)$ denotes the vector of squared singular values of the matrix $A$. This coincides with the standard definition of entanglement entropy for quantum states.

Using the classic approximability results for MPS~\cite{Schuch2008}, we note that large exact entanglement across an MPS cut can obstruct exact low-bond-dimension representations. Approximate representability, however, must be assessed from the decay of the Schmidt coefficients; large exact bond dimension alone does not rule it out.

Sparse coefficient tensors provide a simple sufficient example. If $\mathbf{C}$ is $K$-sparse, with only $K=O(\mathrm{poly}(L))$ nonzero entries, then its MPS bond dimension satisfies $\chi(\mathbf{C}) \le K$. Therefore, our algorithm applies to quantum functions that are also learnable by other approaches whose efficiency relies on coefficient sparsity.

In fact, our algorithm is expected to work for \emph{concentrated-coefficient} functions, which form a broader class than sparse functions. A useful general guideline is that FTCI is efficient when the coefficient tensor does not \emph{spread} too much over the exponentially large basis: this spreading can be quantified by the R\'enyi entropy of the coefficient entries. In particular, we use the following lemma.

\begin{lemma}[Spreading bounds entanglement]\label{lem:ent<concent}
    Given a normalized tensor $\mathbf{C}$ with $L$ legs, for any bipartition $(A:B)$ and any $\alpha\geq0$,
    \begin{equation}
         E^{(\alpha)}_{(A:B)}(\mathbf{C}) \leq S^{(\alpha)}(\mathbf{C}).
    \end{equation}
\end{lemma}

A proof can 
be found in Appendix~\ref{app:entanglement_bound}. Therefore, if the coefficient tensor is concentrated in the sense that $S^{(\alpha)}(\mathbf{C}) = O(\log L)$, then we expect that an MPS with $\chi=O(\mathrm{poly}(L))$ suffices to approximate it. In particular, for $\alpha<1$, the existence of an efficient MPS representation is guaranteed by \jetext{Refs.}~\cite{Schuch2008,verstraete2006matrix}.

\subsection{When is quantum-function learning efficient?}\label{sec:spreading_quantum_functions}

The previous subsection reduced the efficiency question for FTCI to a structural property of the coefficient tensor $\mathbf{C}$: the quantum function is efficiently MPS representable whenever $\mathbf{C}$ is well approximated at polynomial bond dimension. For quantum functions, this leads to a natural physical question.
\begin{center}
\textit{Which properties of the underlying controlled quantum process give rise to a low-bond-dimension coefficient tensor?}
\end{center}

Answering this question for general circuits or Hamiltonian dynamics is difficult, since the map from a quantum process to $\mathbf{C}$ can itself be highly nontrivial. In practice, one can therefore increase the maximum bond dimension used by TCI until either the desired accuracy or the available computational budget is reached. For the family of quantum functions introduced in Sec.~\ref{ss:example_encoding_first}, however, Eq.~\eqref{eq:coefficient_special} makes this relation explicit: for arbitrary pre-processing functions $\phi^{(k)}$, the coefficient tensor $\mathbf{C}$ is obtained by locally restricting the Pauli coordinates of the Heisenberg-evolved observable $O'=W^\dagger O W$ to $I$, $X$, and $Y$. The Pauli spreading of $O'$ therefore directly constrains the tensor-network representation of the resulting function. In particular, we have the following:

\begin{proposition}[Bounded spreading implies efficient MPS representability]\label{prop:pauli_spreading}
Let $0\leq\alpha<1$, and as per Sec.~\ref{ss:example_encoding_first}, consider the quantum function
\begin{equation}\label{eq:spreading_quantum_function}
        f_Q(x)=\bra{0}^{\otimes n}U^{\dagger}(x)OU(x)\ket{0}^{\otimes n},\qquad x\in\mathbb{R}^d,
    \end{equation}
    where
    \begin{equation}
        U(x)=W\bigotimes_{k=1}^n e^{-i\frac{Z^{(k)}}{2}\phi^{(k)}(x)}H^{(k)}
    \end{equation}
for arbitrary pre-processing functions $\phi^{(k)}:\mathbb{R}^d\rightarrow\mathbb{R}$, and where $O=O^{\dagger}\neq0$ satisfies $\norm{O}_{\infty}\leq1$. Let $O':=W^{\dagger}OW$. For $\boldsymbol{\sigma}\in\{0,1,2,3\}^n$, define the full Pauli-coordinate tensor of $O'$ by
    \begin{equation}
        \mathbf{P}_{\boldsymbol{\sigma}}(O')
        :=\frac{1}{2^{n/2}}\Tr\!\left(P_{\boldsymbol{\sigma}}O'\right),
    \end{equation}
Let $\widehat{\mathbf{P}}(O'):=\mathbf{P}(O')/\norm{\mathbf{P}(O')}_2$. If
    \begin{equation}
        S^{(\alpha)}\!\left(\widehat{\mathbf{P}}(O')\right)=O(\log n),
    \end{equation}
   then, for every $\varepsilon>0$, the coefficient 
   tensor $\mathbf{C}$ in Eq.~\eqref{eq:coefficient_special} admits an MPS approximation $\widetilde{\mathbf{C}}$ of bond dimension polynomial in $n$ and $1/\varepsilon$ such that
    \begin{equation}
        \norm{\mathbf{C}-\widetilde{\mathbf{C}}}_2
        \leq\varepsilon\norm{O}_{\mathrm{HS}}.
    \end{equation}
\end{proposition}

\begin{proof}
    By Lemma~\ref{lem:ent<concent}, the assumption implies that, for every bipartition $(A:B)$,
    \begin{equation}
        E^{(\alpha)}_{(A:B)}\!\left(\widehat{\mathbf{P}}(O')\right)
        =O(\log n).
    \end{equation}
    The MPS approximability result of Ref.~\cite{verstraete2006matrix} therefore yields an MPS $\widetilde{\mathbf{P}}$ of bond dimension polynomial in $n$ and $1/\varepsilon$ satisfying
    \begin{equation}
        \norm{\widehat{\mathbf{P}}(O')-\widetilde{\mathbf{P}}}_2
        \leq\varepsilon.
    \end{equation}
    Let $Q=\bigotimes_{k=1}^n Q_k$ denote the local orthogonal projection that retains the Pauli indices $\{0,1,2\}$ and removes index $3$. Equation~\eqref{eq:coefficient_special} then gives
    \begin{equation}
        \mathbf{C}=Q\mathbf{P}(O').
    \end{equation}
    Applying a product of local projections to an MPS does not increase its bond dimension. Moreover, $Q$ is contractive in the Euclidean norm. Therefore, the MPS
    \begin{equation}
        \widetilde{\mathbf{C}}
        :=\norm{\mathbf{P}(O')}_2Q\widetilde{\mathbf{P}}
    \end{equation}
    satisfies
    \begin{equation}
    \begin{split}
        \norm{\mathbf{C}-\widetilde{\mathbf{C}}}_2
        &\leq\norm{\mathbf{P}(O')}_2
        \norm{\widehat{\mathbf{P}}(O')-\widetilde{\mathbf{P}}}_2\\
        &\leq\varepsilon\norm{\mathbf{P}(O')}_2\\
        &=\varepsilon\norm{O}_{\mathrm{HS}}.
    \end{split}
    \end{equation}
\end{proof}

The function family in Eq.~\eqref{eq:spreading_quantum_function} contains BQP characteristic functions as a special case. Set $x\in\{0,1\}^n$ and choose the Boolean pre-processing functions from Eq.~(23), $\phi^{(k)}(x)=\pi x_k$. Up to an irrelevant phase, the input-dependent rotations prepare the Boolean input in the Hadamard basis; the fixed unitary $W$ can include the basis change to the computational basis followed by the BQP decision circuit. In the canonical decision circuit, acceptance is determined by measuring the first output qubit, so we take the one-qubit observable $O=Z^{(1)}$. Thus, approximating $f_Q$ is equivalent to approximating the acceptance probability up to the affine transformation $p_{\mathrm{acc}}(x)=(1-f_Q(x))/2$, when acceptance corresponds to the first output qubit being $|1\rangle$.

For this BQP specialization, the normalized local dictionary is $\mathbf{\Phi}^{(k)}(x)=2^{-1/2}(1,(-1)^{x_k},0)^{\top}$. Let $\mathbf{\Phi}(x):=\bigotimes_{k=1}^n\mathbf{\Phi}^{(k)}(x)$ and $\mathbf{D}:=\mathbf{C}-\widetilde{\mathbf{C}}$. Under the uniform distribution on Boolean inputs,
\begin{equation}
    \mathbb{E}_{x}\!\left[\mathbf{\Phi}(x)\mathbf{\Phi}(x)^{\!\top}\right]
    =G_{\mathrm B}^{\otimes n},\qquad
    G_{\mathrm B}=\operatorname{diag}\!\left(\frac12,\frac12,0\right).
\end{equation}
Moreover, $O=Z^{(1)}$ satisfies $\norm{O}_{\infty}=1$ and $\norm{O}_{\mathrm{HS}}^2=2^n$. Hence Proposition~\ref{prop:pauli_spreading} yields the absolute function-error bound
\begin{equation}
    \begin{split}
        \mathbb{E}_{x\sim\operatorname{Unif}(\{0,1\}^n)}
        \left|f_Q(x)-\widetilde f_Q(x)\right|^2
        &=\mathbf{D}^{\top}G_{\mathrm B}^{\otimes n}\mathbf{D}\\
        &\leq2^{-n}\norm{\mathbf{D}}_2^2\leq\varepsilon^2.
    \end{split}
\end{equation}
Thus, although the general coefficient estimate is naturally measured relative to $\norm{O}_{\mathrm{HS}}$, the canonical single-qubit measurement and the uniform Boolean input distribution cancel this scale exactly. Logarithmic Pauli spreading therefore gives a polynomial-bond-dimension approximation of the complete BQP characteristic function with absolute $L^2$ error at most $\varepsilon$.

Proposition~\ref{prop:pauli_spreading} connects the physical quantum process to its classical TN surrogate for the function family in Eq.~\eqref{eq:spreading_quantum_function}. The spreading of the Heisenberg-evolved observable under $W$ controls the required MPS bond dimension of the coefficient tensor. The quantity $S^{(\alpha)}(\widehat{\mathbf{P}}(O'))$, which measures this spreading in the Pauli basis, is closely related to the \emph{operator stabilizer entropy} (OSE) introduced in Ref.~\cite{dowling2025magic}. Logarithmic spreading thus provides a physical sufficient condition for efficient MPS representability; for the canonical BQP characteristic circuit, it further yields an absolute $L^2$ approximation guarantee for the complete Boolean map $x\mapsto f_Q(x)$ under the uniform distribution on problem instances, rather than a simulation guarantee for a single fixed input.

It is important to emphasize, however, that the condition in Proposition~\ref{prop:pauli_spreading} is sufficient but not necessary. The true bottleneck of FTCI is the MPS entanglement (bond-dimension) of the coefficient tensor. Consequently, FTCI can in principle succeed in regimes where the Pauli coefficients are highly spread (large OSE), provided that they retain a low-entanglement tensor structure. This is distinct from approaches based on \emph{Pauli truncation}, which approximate $O'$ by keeping only polynomially many Pauli strings with the largest weights, relying directly on sparsity or concentration in the Pauli basis.

\begin{figure*}
    \centering
    \includegraphics[width=1.0\linewidth]{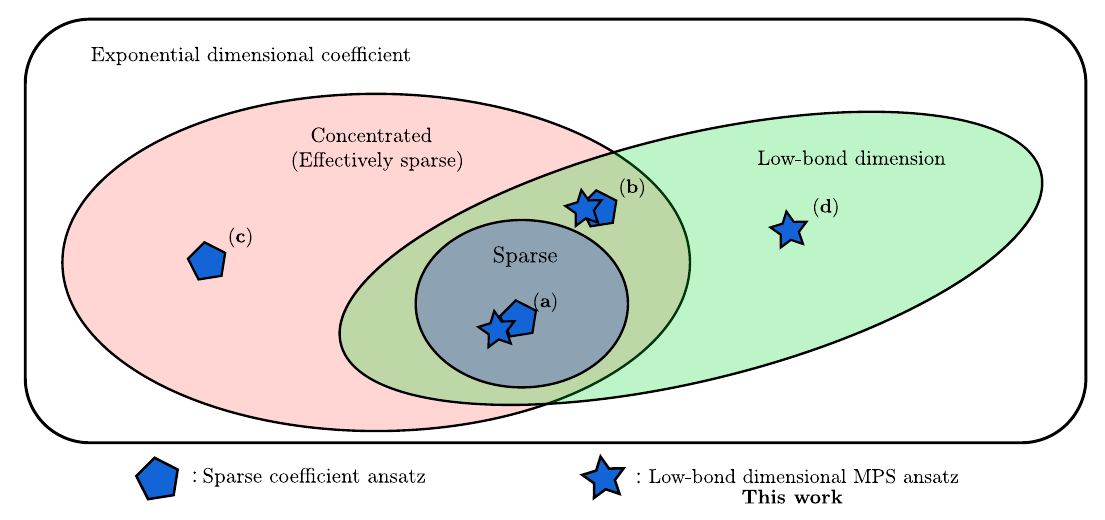}
    \caption{Different types of functions that can be learned efficiently. Here functions are classified by characteristics of coefficient with respect to the same fixed basis functions. (a) Sparse coefficient functions which include low-degree functions. Ref.~\cite{schapire1990strength} works efficiently at this regime. (b) Coefficients that are concentrated and have a low-bond dimensional MPS representation at the same time. Functions that can be learned from random Fourier feature method with efficient MPS based sampling~\cite{Sweke2025potential} lies here. (c) The region of concentrated but high bond-dimensional coefficients. (d) The region where not concentrated (weights of coefficient are evenly distributed) but allows low-bond dimensional representation. Pentagon represents the learning strategy using only manageable (polynomially scaling) number of basis functions, i.e., sparse coefficient ansatz. Star represents the learning algorithm from this work.  (a),(b),(c) regions can be learned with sparse coefficient, meanwhile (b),(c),(d) regions are approximable by low-bond dimensional coefficient.} 
    \label{fig:Bendiagram}
\end{figure*}

A simple extreme example illustrating the separation between spreading and entanglement is a coefficient tensor with uniform distribution. In such a case the global entropy of the coefficient vector can be maximal, while the entanglement across any bipartition remains zero. Identifying physically motivated families of circuits for which such a separation occurs is an interesting direction for future work.

\section{Comparison to existing methods}\label{sec:comparison}

There is a large body of research on learning quantum systems, differing in learning objectives, algorithm type, and data access methods. Here we focus on comparing FTCI with existing methods for classically learning $f_Q(x) = \Tr[\rho(x)O]$ given a classical training set $\{x_i, f_Q(x_i)\}$, potentially together with \jetext{some promises about the structure} of the unknown process.

\subsection{Fourier based classical surrogates of QML models}

The process of \textit{classically surrogating} a variational QML model refers to the process of extracting from a trained variational QML model, a sufficiently accurate classical 
approximation---the classical surrogate---which can then be used for inference~\cite{Schreiber2023surrogate}. As discussed previously, our formulation of 
``learning quantum functions'' is strictly more general than the problem of constructing a classical surrogate for a QML model, including it as a special case. As such, here we briefly review existing methods for \jetext{classical surrogation} (via exploitation of the Fourier spectrum of a quantum model) and compare them to FTCI in this context.

Early work on classical surrogates for variational QML models exploited the well-known 
fact that such models can be represented as finite-dimensional Fourier series~\cite{Schuld2021effect}.  Using this, Ref.~\cite{Schreiber2023surrogate} used a linear-regression approach based on the Fourier basis of the given model, and showed robustness against statistical errors inherited from quantum measurements. However, because their regression uses the full basis-function set, which can be as large as $O((2D+1)^L)$, where $L$ is the dimension of inputs, it suffers from the curse of dimensionality. Hence, it is effective only when the target function has polynomially many basis functions, or one can say when the function is low-dimensional. Our algorithm covers this low-dimensional regime, noticing that \jetext{a low-dimensional function corresponds to a sparse coefficient tensor} in function space spanned by exponentially many basis functions. Moreover, they assume integer, evenly spaced frequencies for the model, unlike our setting where we allow arbitrary real-valued frequencies.

To avoid the curse of dimensionality, Refs.~\cite{landman2022classically,Sweke2025potential,hernicht2025enhancing} adopt the 
\emph{random Fourier features} (RFF) technique. This 
approach samples a moderate number of frequencies from the model's frequency set 
according to an appropriate probability measure, and then uses the sampled frequencies to construct a regression model that \jetext{shows similar performance} to the quantum model. For this method to be efficient, the coefficient tensor of the optimal quantum function needs to be efficiently sampleable, and concentrated in a specific sense. Indeed, \jetext{promising candidates for such coefficient tensors are} those that can be represented via efficient MPS~\cite{Sweke2025potential}. However, in this case one can represent those quantum functions (and their kernel versions) with an efficient TN representation~\cite{shin2025newperspectives,sweke2025kernelbased}. As such, our method is expected to cover these settings. Moreover, the performance of RFF also relies on the alignment between frequency-sampling distribution and the frequency distribution of the target function; in particular, the method functions better when these two distributions are aligned. As such, obtaining good performance is difficult when one does not already know the frequency spectrum of the target function. By contrast, our method does not require such information.

One useful perspective on the relationship between our method and RFF is obtained by comparing RFF and the Nystr\"om method in classical kernel approximation theory~\cite{yang2012nystrom}. The Nystr\"om method approximates a given symmetric positive semi-definite matrix using a 
small number of its columns, and can be viewed as a special case of CUR approximation which is the base algorithm in TCI. Therefore, the relationship between RFF and our method can be understood by leveraging classical results comparing RFF and Nystr\"om-type approximations.

\subsection{Classical shadow based}

We next compare the information returned by FTCI with that available from classical shadows. Classical shadows provide a classical representation of a quantum state $\rho$ that can be used to predict linear properties of the form $\Tr[\rho\,O]$~\cite{huang2020predicting}. The sample complexity required to obtain an accurate shadow depends on the observable $O$ and the measurement scheme. If one can obtain an input-dependent classical shadow $\hat{\rho}(x)$, then it can be used to construct a classical representation of a quantum function via $\Tr[\hat{\rho}(x)O]$.

In Ref.~\cite{huang2022provably}, the authors showed that the classical-shadow-based ansatz
\begin{equation}\label{eq:huang's ansatz}
    \Tr[\hat{\rho}(x)O] \;=\; \Tr\!\left[\frac{1}{N}\sum_{i=1}^N K(x, x_i)\,\rho_T(x_i)\,O\right]
\end{equation}
can be used to predict few-body ground-state properties of parametrized families of smooth Hamiltonians $H(x)$. Here $K(\cdot,\cdot)$ is a chosen kernel function, and $\rho_T(x_i)$ denotes the classical shadow of the state at input $x_i$. Thus, this can be viewed as kernel learning with target values obtained from classical shadows, namely $\Tr[\rho_T(x_i)O]$. This work focuses on ground states of parametrized Hamiltonians and therefore does not fall directly under the notion of quantum functions defined here. Nevertheless, if one constructs a quantum imaginary-time evolution circuit $U_{\mathrm{ITE}}(x)$ that prepares an approximate ground state of $H(x)$, then
\begin{equation}
    f_Q(x)= \bra{0}U^{\dagger}_{\mathrm{ITE}}(x)\,O\,U_{\mathrm{ITE}}(x)\ket{0}
\end{equation}
becomes a ground-state observable function of the type considered in Ref.~\cite{huang2022provably}. This line of work was later improved in Ref.~\cite{lewis2024improved} by using geometrically local, Hamiltonian-informed indicator functions as a feature map together with LASSO-based 
optimization.

In Ref.~\cite{du2025efficient}, the authors learned quantum functions of the form $f_Q(x)=\Tr[\rho(x)O]$, where $\rho(x)=U(x)\rho\,U^{\dagger}(x)$ and $U(x)$ consists of $d$ $R_z(x_i)$ gates and Clifford gates, which is closer to our work than Ref.~\cite{huang2022provably}. They adopt the same kernel ansatz as in Eq.~\eqref{eq:huang's ansatz}, but use a truncated trigonometric kernel $k(x,x')=\Phi^{\top}(x)\Phi(x')$, with feature map
\begin{equation}
    \Phi_{\omega}(x) = \prod_{i=1}^d
    \begin{cases}
        1 & \omega_i=0,\\
        \cos x_i & \omega_i=1,\\
        \sin x_i & \omega_i=-1,
    \end{cases}
\end{equation}
where $\omega\in\{-1,0,1\}^d$ satisfies $\norm{\omega}_0\leq \Lambda$ for a truncation parameter $\Lambda$.

The works discussed above provide efficiency guarantees for smooth quantum functions. At a high level, these approaches and ours can be compared through (i) the choice of feature map (basis functions) and (ii) the coefficient-learning method. Ref.~\cite{huang2022provably} uses a Dirichlet- or shadow-based kernel to learn the coefficients via the induced feature map. Ref.~\cite{du2025efficient} likewise uses a kernel method for coefficient learning, but with a truncated feature map tailored to a subset of the full basis function space. Ref.~\cite{lewis2024improved} uses a geometrically local, Hamiltonian-informed feature map together with LASSO to learn sparse coefficients. In contrast, our algorithm uses the \textit{full} feature map spanning the function space in which the quantum function lies. For coefficient learning, however, we restrict the coefficient tensor to the class of low-bond-dimension MPS and use an active-learning scheme based on TCI. Accordingly, our algorithm is efficient when the coefficient tensor is well approximated by an MPS with small bond dimension, a condition that includes the sparse- and concentrated-coefficient regimes discussed in Sec.~\ref{sec:conditions_for}.

In the context of classical surrogation of variational QML models, Ref.~\cite{Jerbi2024shadows} observed that any variational QML model can be transformed into a so-called flipped model,
\begin{equation}
    f_Q(\theta;x) = \Tr[\rho(\theta)O(x)],
\end{equation}
where one encodes the input into the observable and optimizes the parameters $\theta$ to obtain the trained model. This enables one to collect classical shadows of the trained state $\rho(\theta)$, which are independent of the input $x$; one then constructs a single shadow of a fixed state and uses it to predict outputs for different inputs. We can categorize a flipped model as a quantum function with a fixed state, with all gates before measurement serving as encoding gates.

Finally, a critical difference between classical-shadow approaches and our method is the dependence on observable properties. For classical shadows, the sample complexity typically scales with the shadow norm, which depends on the random measurement scheme and can also reflect quantities such as the operator norm or the locality of the observable~\cite{huang2020predicting}. In contrast, our method does not explicitly depend on these observable-dependent quantities, but is instead governed by the entanglement entropy of the quantum-function coefficient tensor.

\subsection{Classical simulation based}

A final point of comparison is direct classical simulation of the underlying quantum process. When the given quantum system is fully classically simulable, then by definition, one has an efficient way to identify and calculate the function value. Classical tensor network methods~\cite{markov2008simulating}, and recent Pauli propagation-based algorithms~\cite{rudolph2023classicalsurrogate} fall into this category.  
However, classical simulation of quantum functions has quite different settings from our work. First of all, simulation requires the description of the quantum system, such as quantum gates, observables, initial state, and so on. But in our setting we only require the knowledge of encoding generators $\mathcal{G}$, and pre-processing functions. Except for that information, we treat the quantum function as a black box. Second, classical simulation does not require queries from the quantum function, but we need active queries on them. Finally, the efficiency of our algorithm only depends on \jetext{the characteristics of the final quantum function}, not on \jetext{the quantum system's complexity} itself. Therefore, one can think of the case where \jetext{simulation complexity becomes large} during \jetext{the quantum system's evolution}, but \jetext{eventually yields a simple function}, so that learning algorithms based on classical simulations are inefficient, but our algorithm is efficient. This argues that our learning algorithm covers \jetext{a different regime of functions from those accessible to simulation-based algorithms}.

Additionally, here we emphasize one important insight.
In this work we have been \jetext{dealing with TN representations} of quantum \textit{functions}, so it is easy to be confused with TN representation of quantum \textit{states}. However, both are completely different and entanglements are also not directly related to each other, except for special cases. Consequently, our algorithm is fundamentally different from tensor-network–based simulation methods for quantum circuits.

\section{Numerical results}\label{sec:numerics}

Here we present numerical experiments from several different settings, illustrating the breadth of the quantum-function formalism and of our algorithm. We first benchmark FTCI on artificial functions that satisfy our assumptions by construction, and then apply it to quantum functions.
\smallskip

\noindent\textbf{Simulation backend.} For the TCI subroutine, we used the \emph{TensorCrossInterpolation.jl} package~\cite{Yuriel2025learning}, and adopted the $2$-site, full pivot algorithm which requires $O(\chi^2M_{max}^2L)$ samples and $O(\chi^3M_{max}^2L)$ algebra cost, where $\chi$ is \jetext{the prescribed maximum bond dimension} for our MPS ansatz. In the specific package we are using, we set the error-tolerance parameter $\tau$ for the TCI algorithm. Our TCI algorithm would then increase the bond dimension until the relative approximation error becomes lower than $\tau$ or it hits the designated bond dimension cap $\chi$.

For evaluating the quantum function oracle $f_Q$s, we implement exact statevector simulation via the Yao.jl package~\cite{yao_jl}. For larger chains ($N > 16$), we switch to \emph{matrix-product-state} (MPS) based \emph{time-evolving block decimation} (TEBD) via the ITensors.jl package~\cite{itensors_jl}, which scales as $O(n\chi_{\mathrm{sim}}^3)$ per call, where $\chi_{\mathrm{sim}}$ is the MPS bond dimension cap for the simulation (distinct from $\chi$, the bond dimension of our FTCI coefficient tensor). We fix $\chi_{\mathrm{sim}}=128$ and the SVD cutoff at $10^{-12}$.
\smallskip

\noindent \textbf{FTCI details.} For the auxiliary input points $\{a^{(k)}_{j_k}\}_{j_k=1}^{M^{(k)}}$, we sample each $a^{(k)}_{j_k}$ uniformly randomly from $[0,\,2\pi/\gamma^{(k)}+1 := P^{(k)})$, where $\gamma^{(k)}$ is defined in Algorithm~
\ref{alg:quantum_function_ftci}. 
We set $M^{(k)} =2( 2D^{(k)} + 1)$ for all $k$, which satisfies the requirement $M^{(k)}\geq 2D^{(k)}+1$. Random sampling of $a^{(k)}_{j_k}$ in the range $[0,\, P^{(k)})$ follows from Theorem~\ref{thm:error_supression}.

We set the TCI error tolerance $\tau=10^{-8}$ for all
experiments. We denote the maximum bond dimension of the approximated coefficient tensor $\tilde{\mathbf{C}}$ as $\chi(\tilde{\mathbf{C}})$. To evaluate the accuracy of quantum function learning with FTCI, we estimate a relative $L^2$ error by Monte Carlo sampling $M_{\mathrm{test}}$ points $x_i\sim\mathrm{Unif}([0,2\pi]^L)$ according to
\begin{equation}\label{eq:rel_error}
    \hat{\varepsilon}_{\mathrm{rel}}
    = \sqrt{\frac{\sum_{i=1}^{M_{\mathrm{test}}}|f(x_i)-\tilde{f}(x_i)|^2}{\sum_{i=1}^{M_{\mathrm{test}}}|f(x_i)|^2}}.
\end{equation}

\subsection{FTCI on artificial functions}

We first study FTCI on synthetic functions with a prescribed tensor-product Fourier 
feature basis. We use $L$ local feature vectors
\begin{align}
    \mathbf{B}^{(k)}(x_k)= \frac{1}{2}[1~\cos(\omega^{(k)}_1x_k)~\sin(\omega^{(k)}_1x_k)~\cdots \nonumber\\
    ~\cos(\omega^{(k)}_3x_k)~\sin(\omega^{(k)}_3x_k)]^\top,
\end{align}
where the frequencies $\omega^{(k)}_j$ are sampled uniformly from \edit{$[0,1)$}. The resulting function space has dimension $7^L$. For FTCI we use $M^{(k)}=14$ auxiliary points per dictionary, i.e., twice the dictionary dimension $Q^{(k)}=7$.

\subsubsection{Random low bond-dimensional coefficients}

\begin{figure}
    
    \centering
    \includegraphics[width=1.0\linewidth]{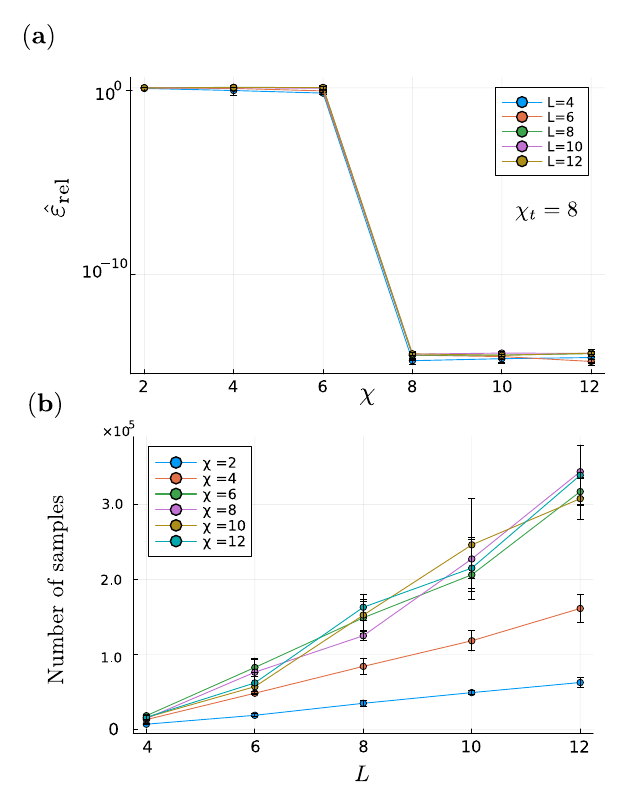}
    \caption{ FTCI on a random small bond-dimensional function. The target function has been generated by randomly generating a coefficient tensor with bond-dimension $\chi_t=8$. All points are averaged values of 5 different random seeds. (a) Relative error vs bond dimension of the approximate function. (b) Number of samples used when TCI is used. }
    \label{fig:Random low chi_t}
\end{figure}

In order to benchmark the functioning of the approach, we generate random target coefficient MPS with bond dimension $\chi_t = 8$, for $L = \{4,6,8,10,12\}$. For bond dimension $\chi$ of our TCI algorithm, we choose $\chi = \{2,4,6,8,10,12\}$. 
All of the resulting functions $\tilde{f}(x)$ had maximum bond dimension $\min\{\chi, \chi_t\}$. Figure~\ref{fig:Random low chi_t} shows the error $\hat{\varepsilon}_{\rm rel}$ and the number of samples used to learn the function. As one can see in Fig.~\ref{fig:Random low chi_t} (a), the error approaches the machine precision when our ansatz has a bond dimension larger than $\chi_t$. In (b), we confirm that the number of samples used does not scale exponentially with the dimension of input ($L$). Meanwhile, the total number of basis functions is $7^L$, and it is around $1.4\times10^{10}$ when $L=12$. This experiment demonstrates clearly that FTCI is effective for exponentially large-dimensional functions with small bond-dimensional coefficients.

\subsubsection{Comparison with truncated trigonometric monomial kernels}

We now compare FTCI with the \emph{truncated trigonometric monomial kernel} (TMK) learning method of Ref.~\cite{du2025efficient}, which uses only a low-degree trigonometric function basis. This experiment is designed to demonstrate a representational difference between the two approaches. TMK is efficient when the target function is well approximated by low-degree trigonometric monomials, whereas FTCI is designed to exploit the low bond-dimensional coefficient tensor. As we noted \jetext{several times}, these two notions are independent: a coefficient tensor can have small MPS bond dimension while its spectral weight is spread over many supports.

We use synthetic target functions of the form
\begin{equation}
    f_C(x)
    =
    \sum_{\omega \in \{0,\pm 1\}^L}
    C_{\omega_1,\ldots,\omega_L}
    \Phi_{\omega}(x),
\end{equation}
where
\begin{equation}
    \Phi_{\omega}(x) = \prod_{i=1}^d
    \begin{cases}
        \frac{1}{\sqrt{2}} & \omega_i=0,\\
        \frac{1}{\sqrt{2}}\cos x_i & \omega_i=1,\\
        \frac{1}{\sqrt{2}}\sin x_i & \omega_i=-1,
    \end{cases}
\end{equation}
The coefficient tensor $C$ is sampled as a random MPS with target bond
dimension $\chi_{t}=8$ and is normalized in Frobenius norm.  Thus
the target is exactly representable by a low-bond-dimensional MPS, while the full basis has dimension $3^L$. We use
$L\in\{4,6,8,10,12\}$ and average over five random seeds.

For FTCI we use the same algorithm as in the previous numerical
experiments, with maximum bond dimension $\chi=12$ and TCI
tolerance $10^{-8}$.  For TMK, we use the estimator
\begin{equation}
    \widehat f_{\Lambda,n}^{\rm TMK}(x)
    =
    \frac{1}{n}
    \sum_{i=1}^n
    f_C(x^{(i)})\,
    \kappa_{\Lambda}(x,x^{(i)}),
    \label{eq:tmk-estimator}
\end{equation}
with uniformly sampled training points $x^{(i)}\in[-\pi,\pi]^L$ and
TMK
\begin{equation}
    \kappa_{\Lambda}(x,x')
    =
    \sum_{\substack{\omega\in\{0,\pm1\}^L\\
                    \|\omega\|_0\leq \Lambda}}
    2^{\|\omega\|_0}
    \Phi_{\omega}(x)\Phi_{\omega}(x') .
    \label{eq:tmk-kernel}
\end{equation}
\edit{We test $\Lambda\in\{1,2,3,4\}$ and $n\in\{50,100,200,500,10^3,2\times10^3,5\times10^3,10^4,2\times10^4,5\times10^4\}$.}  The number of retained TMK features is
\begin{equation}
    M(L,\Lambda)
    =
    \sum_{\ell=0}^{\Lambda}
    {L\choose \ell}2^\ell ,
    \label{eq:tmk-feature-count}
\end{equation}
which is polynomial in $L$ for fixed $\Lambda$.  For example, at
$L=12$, $M(12,3)=2049$, while the full feature has $3^{12}=531441$ entries. All errors are measured on 1000 independent test points with $\hat{\varepsilon}_{\rm rel}$ as per \jetext{Eq.~\eqref{eq:rel_error}}.

\begin{figure*}[t]
    \centering
    \includegraphics[width=0.88\textwidth]{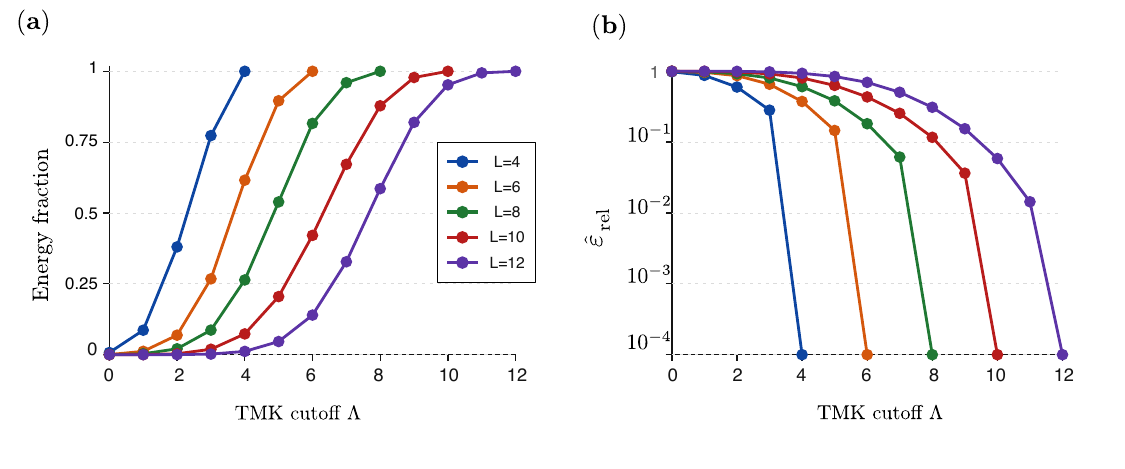}
    \caption{
    \jetext{Degree of spreading} of the low-bond-dimensional target coefficients and the
    best possible TMK truncation error. (a) This shows the
    cumulative coefficient energy retained by monomials with support
    size $\|\omega\|_0\leq \Lambda$.  (b) The right panel shows the
    corresponding best relative $L^2$ error achievable by any estimator
    restricted to the same TMK feature set.  Although the target tensor
    has bond dimension $\chi_{\rm tar}=8$, its coefficient weight is not
    concentrated at small $\|\omega\|_0$.  Consequently, fixed
    low-degree TMK truncations incur a large irreducible bias.
    }
    \label{fig:tmk-ftci-spread}
\end{figure*}

To quantify the intrinsic truncation bias of TMK, we compute \jetext{the quatities}
\begin{equation}
    E_\Lambda^{\rm coef}
    = \frac{\sum_{\|\omega\|_0\leq\Lambda}|C_\omega|^2}
    {\sum_{\omega}|C_\omega|^2},
    \quad
    E_\Lambda^{L^2}
    =
    \frac{
    \sum_{\|\omega\|_0\leq\Lambda}|C_\omega|^2
    \|\Phi_\omega\|_{L^2}^2}
    {\sum_{\omega}|C_\omega|^2
    \|\Phi_\omega\|_{L^2}^2},
    \label{eq:spread-diagnostics}
\end{equation}
and the best possible relative truncation error
\begin{equation}
    \varepsilon_{\rm trunc}^{\star}(\Lambda)
    =
    \sqrt{1-E_\Lambda^{L^2}} .
    \label{eq:best-trunc-error}
\end{equation}

Figure~\ref{fig:tmk-ftci-spread} shows that low bond dimension does not imply concentration on low-support monomials.  For $L=12$ and $\Lambda=3$, the retained $L^2$ energy is only about $4.0\%$, giving $\varepsilon_{\rm trunc}^{\star}\simeq 0.98$.  Even $\Lambda=6$ gives $\varepsilon_{\rm trunc}^{\star}\simeq 0.70$.  Reaching truncation error below $10^{-1}$ requires $\Lambda\simeq 10$, for which $M(12,10)=502769$, already close to the full basis size $3^{12}=531441$. Thus, increasing $\Lambda$ enough to remove the bias also removes the polynomial-size advantage of the truncated TMK feature class.

\begin{figure*}[t]
    \centering
    \includegraphics[width=0.95\textwidth]{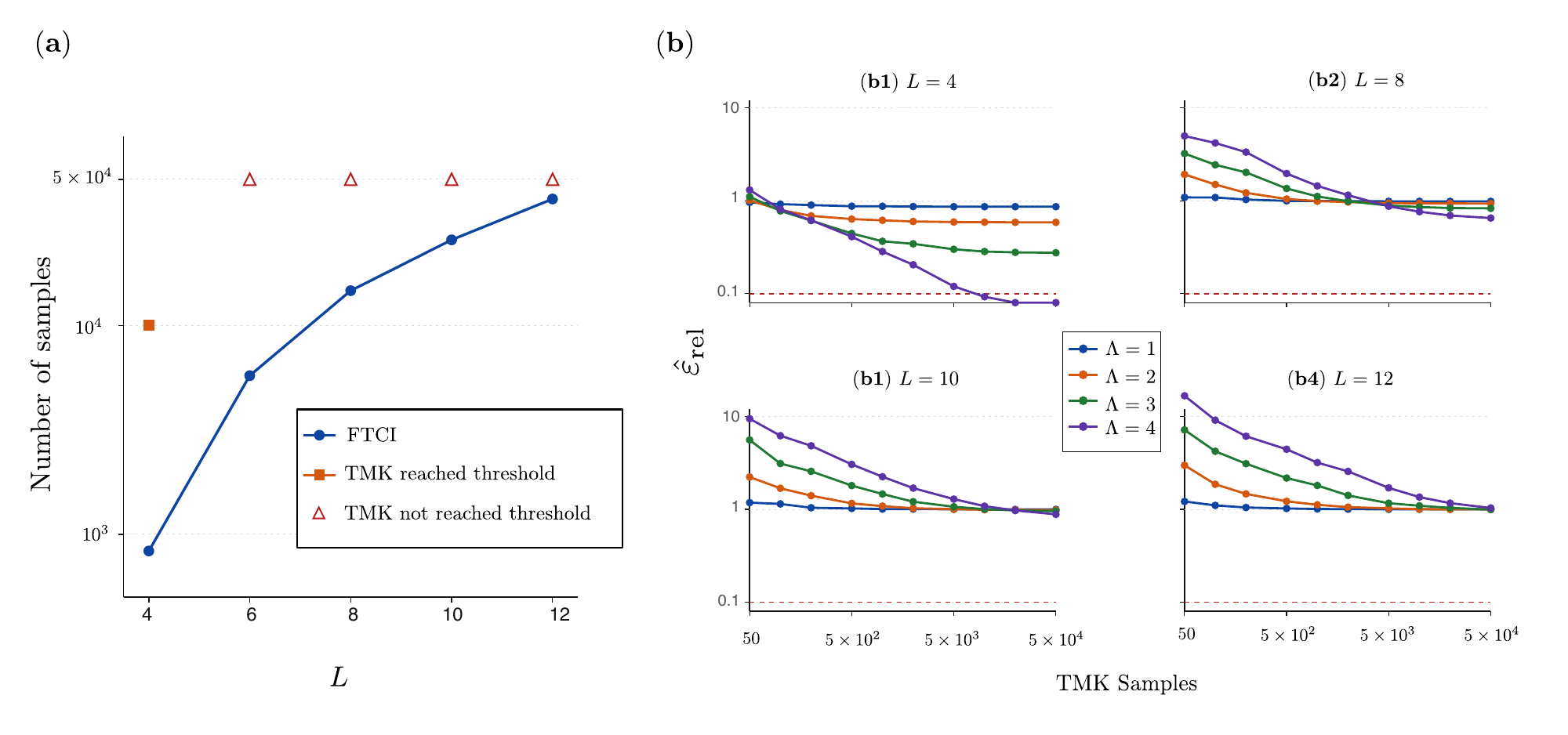}
    \caption{
    Comparison between FTCI and polynomially truncated TMK with
    $n_{\max}=5\times 10^4$ TMK training samples.  (a) Sample counts required to reach $\widehat \varepsilon_{\rm rel}\leq 10^{-1}$. FTCI reaches this threshold for all five random seeds. TMK with $\Lambda\leq 4$ reaches the threshold only for the smallest system size $L=4$, and fails to reach it for $L\geq 6$. Red triangles mark
    TMK failures at the largest tested sample size and should be interpreted as lower bounds, not as successful sample counts.
    (b1)--(b4) TMK learning curves for $L=4,8,10,12$. Increasing the number of TMK samples up to $5\times 10^4$ does not remove the large-$L$ error floor for any $\Lambda\leq 4$, consistent with the truncation-bias diagnostic in Fig.~\ref{fig:tmk-ftci-spread}.}
    \label{fig:tmk-ftci-threshold-multil}
\end{figure*}

The threshold comparison is shown in Fig.~\ref{fig:tmk-ftci-threshold-multil}(a). FTCI achieves essentially exact reconstruction, with mean relative errors of order $10^{-15}$ for all tested system sizes. The corresponding mean numbers of distinct oracle queries are
\begin{equation}
\begin{array}{c|c}
    L & \text{FTCI samples} \\
    \hline
    4  & 8.32\times 10^2  \\
    6  & 5.75\times 10^3 \\
    8  & 1.47\times 10^4 \\
    10 & 2.57\times 10^4 \\
    12 & 4.03\times 10^4
\end{array}
\label{eq:ftci-sample-table}
\end{equation}
In contrast, the polynomially truncated TMK models reach the threshold only at $L=4$, where the choice $\Lambda=4$ already contains the full local-support basis.  The mean TMK sample count at this point is $10^4$.  For every $L\geq 6$, TMK with $\Lambda\leq 4$ does not reach $10^{-1}$ within $5\times 10^4$ samples. This failure is not primarily a finite-sample effect.  The truncation diagnostic in Fig.~\ref{fig:tmk-ftci-spread} shows that, for these low values of $\Lambda$, the best approximation within the TMK model class is already far above the desired error threshold.

Figures~\ref{fig:tmk-ftci-threshold-multil}(b1)---(b4) show the TMK learning curves for representative system sizes. At $L=4$, $\Lambda=4$ contains the full support set and the error decreases below $10^{-1}$, although it still requires substantially more samples than FTCI. For $L=8,10,12$, however, all tested cutoffs $\Lambda\leq 4$ remain far above the $10^{-1}$ threshold even at $5\times 10^4$ samples. This behavior is consistent with the truncation analysis: additional samples can reduce estimation variance, but they cannot recover coefficient weight outside the retained support set $\|\omega\|_0\leq\Lambda$.

These results support the central distinction between the two learning principles.  TMK imposes a low-degree or low-support prior on the trigonometric expansion, while FTCI imposes a low-bond-dimensional tensor prior on the coefficient tensor. The present synthetic target belongs to the latter class but not to the former. Consequently, FTCI learns the function accurately from direct function queries, whereas polynomially truncated TMK suffers from an irreducible approximation bias. This demonstrates that FTCI can access a class of functions that is not efficiently captured by low-degree trigonometric monomial truncation, even when both methods use the same local Fourier dictionary.
Here, we emphasize that we are \textit{not} claiming our FTCI algorithm is better than the TMK algorithm. Instead, we aim here to demonstrate clearly the differences between sparsity-based algorithms like TMK and ours. In particular, we simply illustrate here that there exist functions that reside in the region (d) in Fig.~\ref{fig:Bendiagram}, and for this region, our algorithm is preferable.

\subsubsection{Fast decaying Fourier series}

We next examine a function whose Fourier coefficients decay rather than vanish abruptly.
\begin{figure}
    
    \centering
    \includegraphics[width=1.0\linewidth]{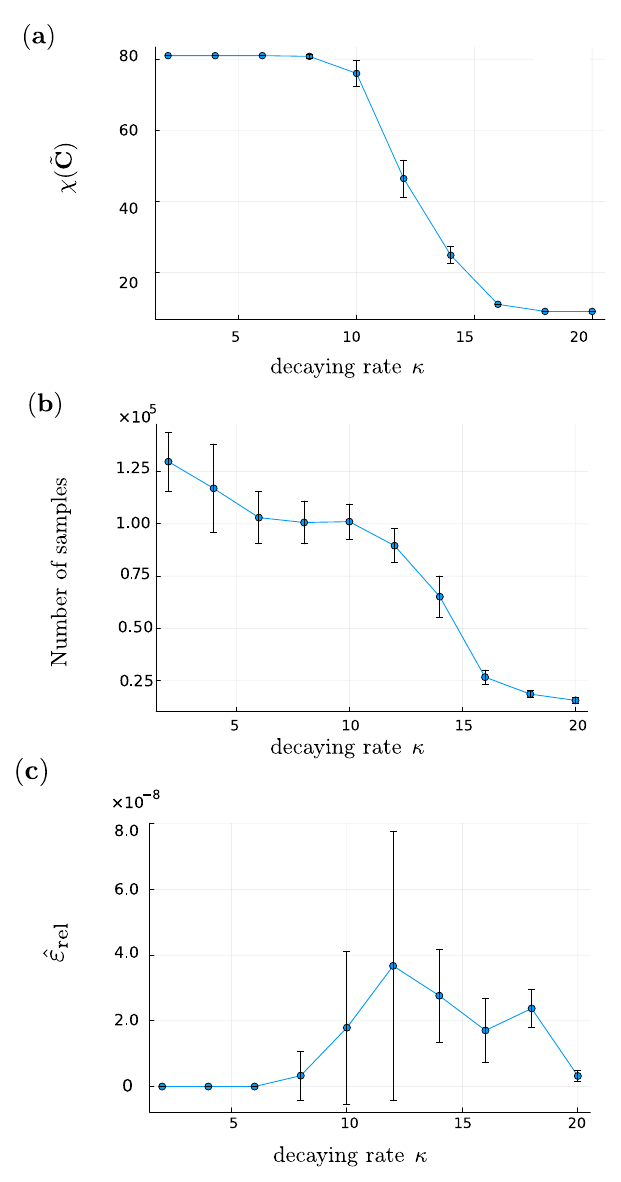}
    \caption{ FTCI on decaying-coefficient function. Target function's coefficient has a power-law scaling with decaying rate $\kappa$. (a) The maximum bond dimension of $\mathbf{\tilde{C}}$ with respect to $\kappa$.  (b) The number of samples used. (c) Relative function error.}
    \label{fig:Decaying}
\end{figure}
Following the discussion in Sec.~\ref{sec:conditions_for}, we expect that \emph{concentrated} coefficient tensors are not only favorable for \textit{sparse-ansatz} based learning algorithms, but also for the FTCI algorithm.

To illustrate this, we consider an $L$-variate Fourier series whose frequency vectors satisfy $\omega\in\{-1,0,1\}^L$. We generate coefficients that decay with the $\ell_1$-norm of the frequency vector
\begin{equation}
    c_\omega \sim \frac{1}{\|\omega\|_1^\kappa}\,u,
    \qquad u \sim \mathrm{Unif}([-1,1]),
\end{equation}
where $c_\omega$ denotes the coefficient associated with the basis function of frequency $\omega$. 
\jetext{
For $\omega=0$, we set it to zero.}  In this construction, increasing $\kappa$ concentrates the spectral weight on low-frequency (low-degree) components.

For FTCI, we use $L$ local function dictionaries
\begin{equation}
    \mathbf{B}^{(k)}(x_k)=
    \begin{pmatrix}
        1\\
        \cos x_k\\
        \sin x_k
    \end{pmatrix},
\end{equation}
which span the target function space induced by frequencies in $\{-1,0,1\}$. We fix $L=8$ and sweep $\kappa\in\{2,4,6,\ldots,20\}$. In the TCI routine we set the maximum bond dimension to $\chi_{\max}=81$, which equals the largest possible MPS bond dimension for $L=8$ with local dimension $3$. This ensures that the approximation error is controlled by the target tolerance $\tau$ rather than by an artificially tight bond-dimension cap we impose.

As shown in \jetext{Fig.~\ref{fig:Decaying}(a)}, increasing the decay rate $\kappa$ indeed reduces the bond dimension required to reach the prescribed tolerance $\tau$. Moreover, the total number of function evaluations (samples) required by FTCI decreases accordingly \jetext{as expected, as shown in panel (b)}. Finally, panel (c) shows the corresponding relative function errors, which remain small over the entire range of $\kappa$.

\subsection{QML model: Quantum convolutional neural networks}

\begin{figure*}   
    \centering
    \includegraphics[width=1.0\linewidth]{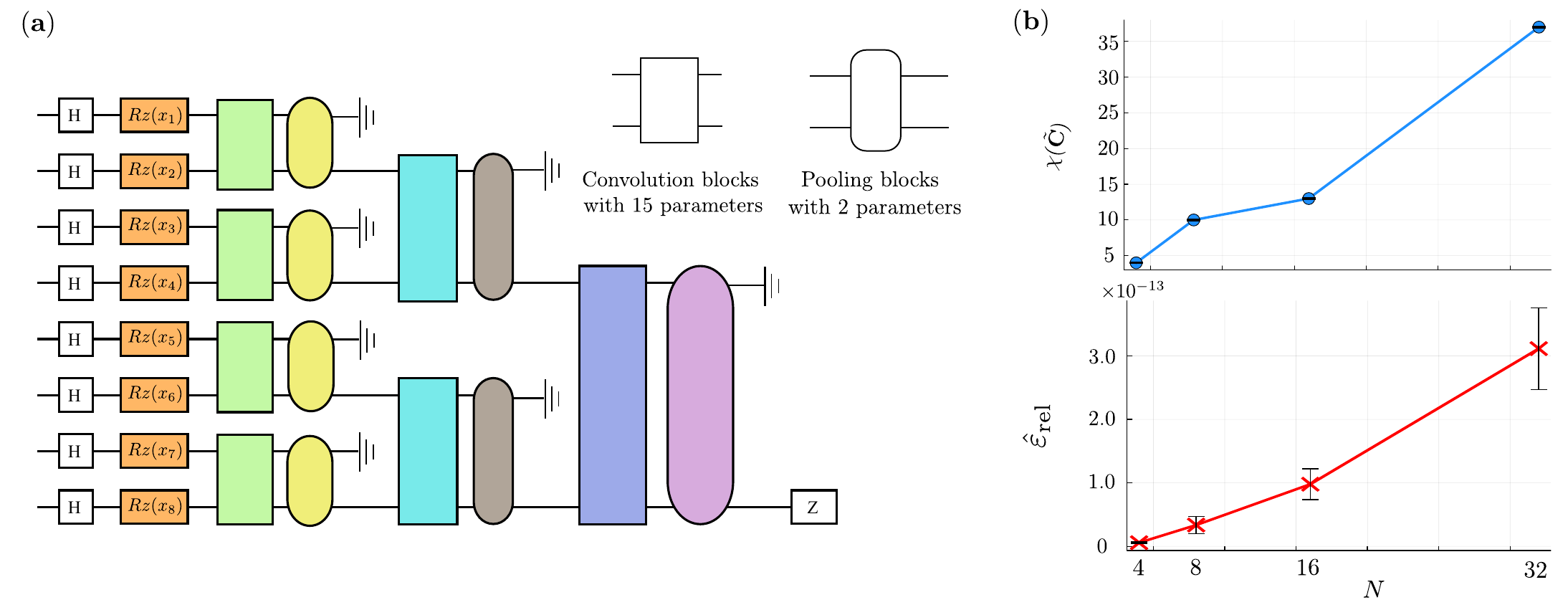}
    \caption{Learning \emph{quantum convolutional neural network} (QCNN) models. (a) Quantum circuit for $N=8$ QCNN model. $N-$bit input is encoded by single-qubit $Z$-rotation unitaries followed by $\ket{+}$ state \je{vectors}. Convolutional 
    layers are followed and $Z$ expectation value is measured. The quantum gates with the same color share the same parameters. (b) The maximum bond dimension of approximating functions' coefficients, and corresponding relative errors. }
    \label{fig:QCNN}
\end{figure*}

We now test FTCI on a quantum convolutional neural network, using its measured output as the target function.
We illustrate here the application of \textit{classical surrogation} of a variational QML model, using a translationally invariant \emph{quantum convolutional neural network} (QCNN). Input-dependent unitaries are given as single-qubit $Z$ rotations that depend on each element of the input vector. After input-dependent unitaries, QCNN layers follow. In this example, we set the number of qubits $N$ equal to the dimension of the input $d$. \jetext{With this, each layer} consists of $2$-qubit convolution blocks using $15$ parameters, followed by pooling blocks using $2$ parameters. After pooling blocks, we discard one of the qubits; thus, we have 
\jetext{$\log_2 N$ layers} until only one qubit remains. See Fig.~\ref{fig:QCNN} (a) for the $d=N=8$ case. For the circuit construction, we used the structure introduced in Ref.~\cite{lee2025optimizing}, Fig.~5. All blocks within the same convolutional/pooling layer share the same parameters. For the output, we measure the single qubit $Z$ operator. 

The quantum function derived from this QCNN has simple function dictionaries,
\begin{equation}
    \mathbf{\Phi}^{(k)}(x_k)=
    \begin{pmatrix}
        1\\
        \cos x_k\\
        \sin x_k
    \end{pmatrix}.
\end{equation} We constructed $N\in\{4,8,16,32\}$ models and \jetext{learned} the functions using these function dictionaries. For quantum-function evaluations, we used the tensor network method with a \jetext{maximum} bond dimension truncated to $64$, for numerical efficiency. Note that this bond dimension for the simulation of \jetext{the} QCNN circuit is \textit{different} from the bond dimension of our FTCI algorithm or of the ansatz. 

\jetext{Function-learning results for five randomly chosen parameter sets are depicted in Fig.}~\ref{fig:QCNN}\jetext{(b), which shows} the relative RMSE and the corresponding $\chi(\tilde{\mathbf{C}})$. As one can see, with only linearly increasing bond dimensions (\jetext{and hence} polynomially increasing computational time with respect to the number of qubits), we could achieve error rates around $10^{-13}$.

\subsection{Global observable for shallow circuit states}

Given a quantum system, often one is concerned with extracting the expectation values of Pauli strings.  In this subsection, we demonstrate how one can learn the values of Pauli string expectations from shallow circuit states \jetext{using FTCI}.
We consider an $N$-qubit quantum state vector $\ket{\psi}$. For each qubit $k\in[N]$, define the single-qubit rotation
\begin{equation}
    R_{n}^{(k)}(x_k)
    := \exp\!\left(-i\,\frac{X^{(k)}+Y^{(k)}+Z^{(k)}}{\sqrt{3}}\,x_k\right),
\end{equation}
where the superscript $(k)$ labels the qubit. For a subset $S\subseteq [N]$, define the product unitary
\begin{equation}
    R_n^{S}(x)
    := \bigotimes_{k\in S} R_n^{(k)}(x_k),
\end{equation}
and the Pauli-$Z$ string on $S$ by $Z^{S}:=\bigotimes_{k\in S} Z^{(k)}$. Our target quantum function is
\begin{align}\label{eq:observable quantumfunction}
    f_Q(x;S,\ket{\psi})
    := \bra{\psi}\,\bigl(R_n^{S}(x)\bigr)^{\dagger}\, Z^{S}\, R_n^{S}(x)\,\ket{\psi}.
\end{align}
For a single qubit (suppressing the qubit index $k$ for clarity), let
\begin{equation}
\begin{split}
    O(x) &:= R_n^{\dagger}(x)\, Z\, R_n(x) \\
    &= \left( \frac{2\sin^2(x)}{3} - \frac{\sin(2x)}{\sqrt{3}} \right)X
     + \left( \frac{2\sin^2(x)}{3} +  \frac{\sin(2x)}{\sqrt{3}} \right)Y
     \\&+ \left( 1 - \frac{4\sin^2(x)}{3} \right)Z,
\end{split}
\end{equation}
so that
\begin{align}
    O(0)=Z,\qquad O(2\pi/3)=X,\qquad O(4\pi/3)=Y.
\end{align}
Therefore, $f_Q(x;S,\ket{\psi}) : \mathbb{R}^{|S|}\rightarrow[-1,1]$ encodes the expectation values of all $3^{|S|}$ Pauli strings supported on $S$,
\begin{equation}
    \Tr\!\left[\ketbra{\psi}{\psi}\,P^{S}\right],
    \qquad P^{S}\in\{X,Y,Z\}^{\otimes |S|},
\end{equation}
by mapping
\begin{equation}
    x_k = 0 \mapsto Z^{(k)},\quad
    x_k = 2\pi/3 \mapsto X^{(k)},\quad
    x_k = 4\pi/3 \mapsto Y^{(k)},
\end{equation}
with
\begin{equation}
    f_Q(x_1\ldots x_{|S|}) = \Tr\left[\ketbra{\psi}{\psi}P_{x_1, \ldots , x_{|S|}}\right]
\end{equation}
For example, let $N = 4$, and $S=\{1,3,4\}$. Then,
\begin{equation}
    f_Q(0,\,2\pi/3,\,4\pi/3;S,\ket{\psi}) = \Tr[\ketbra{\psi}{\psi}Z^{(1)}I^{(2)}X^{(3)}Y^{(4)}].
\end{equation}
By learning this $f_Q$, one can classically predict any $|S|$-local Hermitian observable, up to additional numerical error due to post-processing (e.g., linear combinations of estimated values).

This quantum function has a structure similar to the example in Sec.~\ref{ss:example_encoding_first}. Following the analysis in Ref.~\cite{shin2024dequantizing} again, the bond dimension of the coefficient tensor is bounded by the bond dimension of a \emph{matrix-product-operator} (MPO) 
representation of $\ketbra{\psi}{\psi}$. Consequently, our method performs well for weakly entangled states, such as shallow-circuit states. For instance, when the circuit depth scales as $O(\log N)$, the MPO bond dimension (and hence the coefficient-tensor bond dimension) scales at most polynomially in $N$.

\begin{figure*}
    
    \centering
    \includegraphics[width=1.0\linewidth]{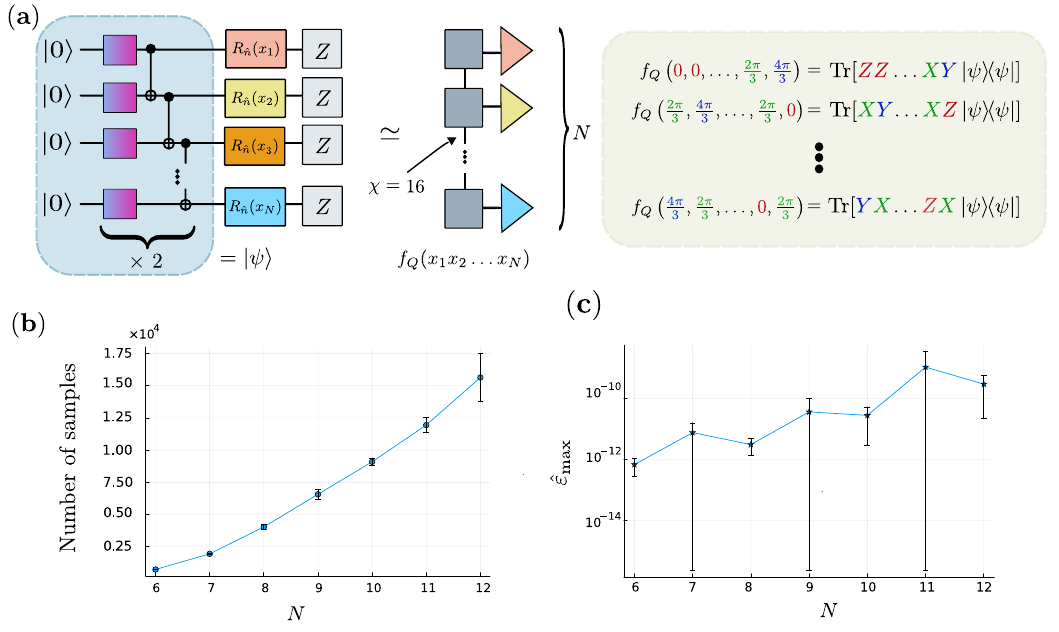}
    \caption{Quantum function that encodes global observable expectation values. (a) Quantum circuit that generates quantum function. Constant depth-$2$ circuit is constructed with random single-qubit gates and CNOT gates. Input-dependent unitaries are applied just before the $Z$ observable measurement. The values of this quantum function correspond to expectation values for Pauli strings when $x_k \in\{0,2\pi/3,4\pi/3\}$. (b) The number of samples used. (c) The maximum relative absolute error. They are estimated over $1000$ random Pauli strings. The maximum bond dimension of coefficient tensors were $16$, regardless of their size. }
    \label{fig:Global observable}
\end{figure*}

As a demonstration, for each $N \in \{6,7,\ldots,12\}$ we generated $5$ random shallow-circuit state vectors $\ket{\psi}$ using a depth-two circuit consisting of two layers of random 
single-qubit gates interleaved with nearest-neighbor CNOT gates, as shown in Fig.~\ref{fig:Global observable}(a). In this setting, the resulting density operator $\ketbra{\psi}{\psi}$ has MPO bond dimension at most $16$~\footnote{This bound follows from the standard MPO representation of a depth-$d$ nearest-neighbor two-qubit circuit, for which the operator Schmidt rank across any cut is bounded by $4^{d}$; here $d=2$.}. We learn the function in Eq.~\eqref{eq:observable quantumfunction} with $S=[N]$. The corresponding function dictionaries are
\begin{equation}
    \mathbf{\Phi}^{(k)}(x_k)=
    \edit{\begin{pmatrix}
        1\\
        \cos 2x_k\\
        \sin 2x_k
    \end{pmatrix}.}
\end{equation}
For the auxiliary points, we choose the deterministic grid $a \in \{0,\,2\pi/3,\,4\pi/3\}^{N}$ rather than sampling randomly. This choice (i) yields nonsingular design matrices $V^{(k)}$, and (ii) corresponds directly to measuring expectation values of Pauli strings. 

Our algorithm learns $f_Q(x;[N],\ket{\psi})$ with $\chi(\tilde{\mathbf{C}})=16$ for \textit{all} $N$, achieving the accuracies shown in Fig.~\ref{fig:Global observable}(c), which are nearly perfect. Here we define
\begin{equation}
    \hat{\varepsilon}_{\max} := \max_i \left|\frac{\Tr\!\left[P_{a_i}\ket{\psi}\bra{\psi}\right] - \tilde{f}(a_i)}{\Tr\!\left[P_{a_i}\ket{\psi}\bra{\psi}\right]}\right|
\end{equation}
as the estimated maximum relative error, where $M=1000$ Pauli strings (i.e., $a_i$'s) are sampled uniformly at random from the total set of $3^{N}$ strings. In Fig.~\ref{fig:Global observable}(b), we also observe that the required number of samples does not scale exponentially with the system size.

One standard approach to classically estimate observable expectation values from a given quantum state is the classical shadow formalism. However, global observables supported on $O(N)$ qubits do not admit an efficient shadow reconstruction~\cite{huang2020predicting}. In contrast, our learning algorithm was able to handle such broadly supported observables, since its complexity bottleneck is the bond dimension of the given quantum state rather than the observable support.

We stress that \jetext{there are other existing approaches} for tomography of low–bond-dimension quantum states, such as approaches based on \emph{tensor cross interpolation}  (TCI)~\cite{lidiak2022quantumstatetomographytensor} and MPS 
tomography~\cite{Cramer-NatComm-2010,Efficient,Lanyon2017}, which could 
also have been used effectively in this setting.
Our example should therefore be viewed as a demonstration of our 
quantum function scheme, providing an alternative route to expectation-value prediction that can be compared with these existing methods. We note, however, that full tomography recovers strictly more information than expectation-value prediction, at a correspondingly higher cost.

\subsection{Hamiltonian dynamics}\label{sec:ham_dynamics}

\begin{figure*}[t]
\centering
\includegraphics[width=\textwidth]{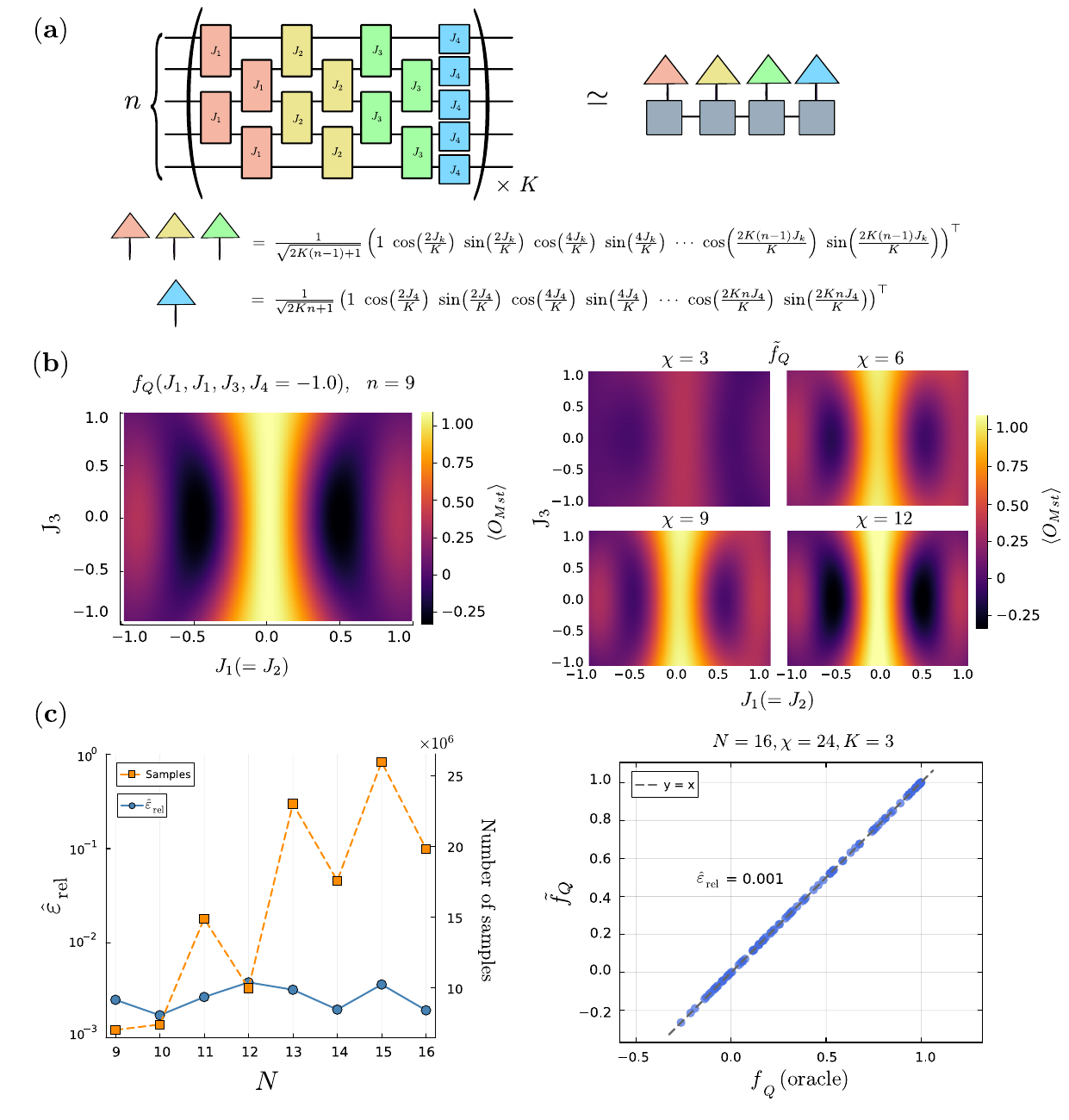}
\caption{Hamiltonian dynamics simulation learning. (a)~$K$-step Trotterized circuit for simulating Hamiltonian dynamics of the Heisenberg chain,
$H(J_1,J_2,J_3,J_4) = \sum_{i=1}^{n-1}(J_1 X_iX_{i+1}
  + J_2 Y_iY_{i+1} + J_3 Z_iZ_{i+1})
  + \sum_{i=1}^{n} J_4 Z_i$.
We denote input-dependent unitaries with the same color when they depend on the same parameter. We use 4 function dictionaries, one for each parameter $J_k$. (b)~Comparison between true quantum function value versus approximated ones with designated bond dimensions. We set $J_1=J_2$, and $J_4=-1.0$. We plotted $200\times 200$ grid points from $(J_1,J_3)\in[-1,1]\times[-1,1]$. (c)~Scalability of FTCI with $\chi=24$ and $K=3$. 
\textit{Left}: relative error $\hat{\varepsilon}_{\mathrm{rel}}$ and the number of oracle samples versus chain length $n$ for $n\in\{9,10,\dots,16\}$.
\textit{Right}: oracle versus FTCI prediction for 100 random test points at $n=16$, confirming accurate approximation across the parameter space.}
\label{fig:ham}
\end{figure*}

\begin{figure*}
    \centering
    \includegraphics[width=1.0\linewidth]{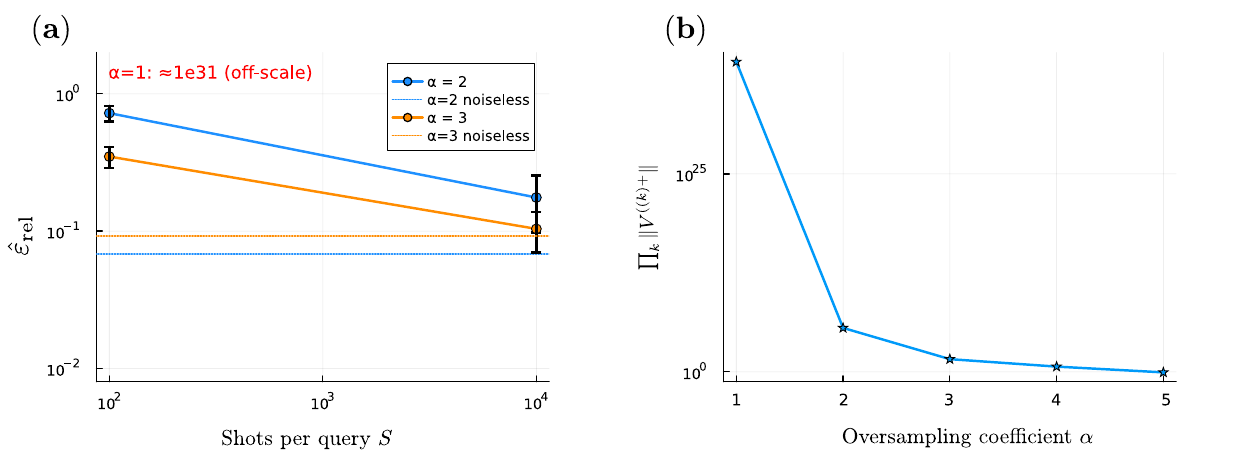} 
    \caption{FTCI with a noisy oracle for the Heisenberg chain ($n=9$, $K=3$, $\chi=12$). (a) Relative error differences among different oversampling coefficients $\alpha \in\{1,2,3\},$ showing the necessity of oversampling. The dotted lines are the noiseless references. (b) 
    Inversion-error propagation factor $\prod_k \|V^{(k)+}\|$ versus $\alpha$. The norm of the product of inversion matrices is ${\approx}1.5\times10^{39}$ for the square design $\alpha=1$, falling steeply with oversampling and reaching $8.487\times 10^{-1}$ by $\alpha=5$.}\label{fig:noisy}
\end{figure*}

We turn to a Hamiltonian example in which the input parameters control the couplings and hence the measured dynamics. We consider a time-independent Hamiltonian $H(J)$ parametrized by $J = (J_1,\dots,J_h)\in\mathbb{R}^{h}$. Our goal is to learn the quantum function
\begin{equation}\label{eq:fQ_ham}
  f_Q(J) := \langle\psi|\,
    U^{\dagger}(J;K,t)\, O\, U(J;K,t)\,|\psi\rangle,
\end{equation}
where $O$ is a fixed observable and
\begin{equation}\label{eq:trotter}
  U(J;K,t) := \biggl(\prod_{b=1}^{B}
    e^{-i H^{(b)}(J)\,t/K}\biggr)^{\!K}.
\end{equation}
denotes a first-order approximation of $e^{-iH(J)t}$ with $K$ Trotter steps.
Here \begin{equation}
H(J)=\sum_{b=1}^{B}H^{(b)}(J) 
\end{equation}
is a decomposition into blocks whose terms mutually commute within each block, so that each factor $e^{-iH^{(b)}(J)\,t/K}$ can be implemented as a layer of commuting one- and two-qubit gates.

In our numerical experiment, we consider the $n$-qubit anisotropic Heisenberg chain with a longitudinal field,
\begin{equation}\label{eq:heisenberg}\begin{split}
  H(J_1,J_2,J_3,J_4) = &\sum_{i=1}^{n-1}
    \bigl(J_1 X_i X_{i+1} + J_2 Y_i Y_{i+1} + J_3 Z_i Z_{i+1}\bigr)
    \\&+ \sum_{i=1}^{n} J_4\, Z_i.
    \end{split}
\end{equation}
We fix the evolution time to $t=1$, and split $H$ into 
$B=7$ commuting blocks by separating even and odd blocks as
\begin{align}\label{eq:trotter_blocks}
  H^{(1)} &:= J_1 \sum_{i\;\mathrm{even}} X_i X_{i+1},\,
  H^{(2)} := J_1 \sum_{i\;\mathrm{odd}}  X_i X_{i+1},  \\
  H^{(3)} &:= J_2 \sum_{i\;\mathrm{even}} Y_i Y_{i+1},\,
  H^{(4)} := J_2 \sum_{i\;\mathrm{odd}}  Y_i Y_{i+1},  \\
  H^{(5)} &:= J_3 \sum_{i\;\mathrm{even}} Z_i Z_{i+1},\,
  H^{(6)} := J_3 \sum_{i\;\mathrm{odd}}  Z_i Z_{i+1},\\
  H^{(7)} &:= J_4 \sum_{i=1}^{n} Z_i.
\end{align}
One Trotter step is then implemented as 
$\prod_{b=1}^{7}e^{-iH^{(b)}/K}$, and the full circuit 
is the $K$-fold repetition of this step. We take the initial state 
to be the N\'{e}el state vector
\begin{equation}\label{eq:neel}
  |\psi\rangle = |1,0,1,0,\cdots\rangle,
\end{equation} 
and measure the staggered magnetization operator,
\begin{equation}\label{eq:O_Mst}
  O_{\rm Mst} := \frac{1}{n}\sum_{i=1}^{n}(-1)^{i}\,Z_i.
\end{equation}
Each term in Eq.~\eqref{eq:trotter_blocks} is a Pauli string with eigenvalues $\pm 1$. Since we use $t=1$, the scaled generator $(1/K)P$ has eigenvalues $\pm 1/K$. Consequently, every gate that depends on a parameter $J_\alpha$ contributes basis functions $\cos\!\bigl(\tfrac{2}{K}J_\alpha\bigr)$, $\sin\!\bigl(\tfrac{2}{K}J_\alpha\bigr)$. Merging all such dictionaries, we obtain four local function dictionaries whose frequencies are integer multiples of $2/K$, up to a maximum $m_\alpha\cdot 2/K$. For the interaction parameters $\alpha\in\{1,2,3\}$, there are
$(n-1)$ two-qubit interaction terms per Trotter step and $K$ steps, and hence
\begin{equation}\label{eq:m_bond}
  m_1 = m_2 = m_3 = K(n-1).
\end{equation}
For the longitudinal field parameter $J_4$, there are $n$ single-qubit terms per step and $K$ steps, which means that
\begin{equation}\label{eq:m_field}
  m_4 = Kn.
\end{equation}
Therefore, we have four function dictionaries with frequency sets
\begin{equation}\label{eq:freq_sets}\begin{split}
  &\Omega_\alpha = \biggl\{\frac{2r}{K}:\;r=0,1,\dots,m_\alpha\biggr\}
  \\
  &m_\alpha = \begin{cases}
    K(n-1), & \alpha\in\{1,2,3\},\\
    Kn,     & \alpha=4.
  \end{cases}
  \end{split}
\end{equation}
We apply FTCI to learn $f_Q(J)$ using function
dictionaries constructed with the above frequency sets, and it provides us a length-4 TN representation with local dimensions $d_\alpha = 2m_\alpha + 1$.

In Fig.~\ref{fig:ham}\jetext{(b)}, we depict the $K=3$ and $n=9$ case with bond-dimension settings $\chi\in\{3,6,9,12\}$. We plot the true quantum function values and approximated ones in the setting $J_1=J_2\in[-1,1]$, $J_3\in[-1,1]$, and $J_4=-1$. In other words, we consider the Heisenberg $XXZ$ model
with a longitudinal field. One can see that increasing the bond dimension improves the approximation accuracy, and the relatively small bond dimension $\chi=12$ was enough to reproduce the phase diagram with a relative RMSE of \edit{$0.070$}, compared to the possible maximum bond dimension for the coefficient tensor, \edit{$49^2 = 2401$}.

Next, we select $K=3$, $\chi=24$, and sweep the chain length over $n\in[9,16]$. For each $n$, we evaluate $\hat{\varepsilon}_{\mathrm{rel}}$ on 500 uniformly random test points drawn from $[-1,1]^4$. The results are summarized in Fig.~\ref{fig:ham}\jetext{(c)}. In Fig.~\ref{fig:ham}\jetext{(c, left panel)}, we observe that the relative error remains below $5\%$ across the entire range $n=9$--$16$. While none of the runs could achieve the prescribed error tolerance $\tau$ for the TCI step, we were able to learn the function with \edit{relative errors below $0.4\%$}. The oracle sample count, shown on the secondary axis, grows moderately with $n$; this growth is dominated by the increasing local dimension $2K(n-1)+1$, which controls the TCI grid size. In Fig.~\ref{fig:ham}\jetext{(c, right panel)}, we show a direct comparison between the oracle values $f_Q(J)$ and the FTCI predictions $\tilde{f}_Q(J)$ for 100 uniformly random test points at the largest chain length $n=16$. The data points cluster tightly around the $y=x$ diagonal, confirming that the learned approximation accurately captures the quantum function across the parameter space. These results show that learning quantum functions from Hamiltonian dynamics using FTCI can scale gracefully with system size: The number of oracle queries remains tractable.

\subsection{Robustness to shot noise}\label{sec:noisy_oracle}

Finally, we address the issue of robustness to noise. The experiments presented so far assumed the ability to make \textit{exact} queries to the unknown quantum function. However, as we have discussed earlier, in practice, shot noise (at least) always leads to \textit{approximate function queries.}
In particular, each oracle call returns an expectation value estimated from a finite number of measurement shots $S$, so that every queried entry of the virtual tensor carries statistical noise of magnitude $O(1/\sqrt{S})$. Our task therefore becomes a noisy TCI. However, as noted in subsection ~\ref{ss:Error_analysis_for_quantum}, errors from TCI algorithms, including noise-oriented ones, can be mitigated by oversampling. In other words, by choosing the number of auxiliary grid points as $M^{(k)}=\alpha(2D^{(k)}+1)$, with $\alpha > 1$.

To demonstrate this, we use the Hamiltonian-dynamics setting of Sec.~\ref{sec:ham_dynamics}. We model an $S$-shot oracle by returning $f_Q(a) + \varepsilon$ with $\varepsilon \sim \mathcal{N}(0, 1/S)$ for each query. We use the chain length $n=9$, $K=3$ Trotter steps, bond-dimension cap $\chi=12$. 
First, we used oversampling constant $\alpha \in\{1,2,3\}$, and sweep $S \in \{10^2, 10^4\}$ over $3$ random seeds. In Fig.~\ref{fig:noisy} (a), one can observe that the relative error decreases as the shot count $S$ increases, approaching the noiseless reference value. As one can see, the oversampling ($\alpha > 1$) is essential here. When there is no oversampling ($\alpha=1$), the error blows up. Meanwhile, we could achieve decent performance for $\alpha = 2$ and $3$ in the $S = 10^4$ regime. In Fig.~\ref{fig:noisy} (b), we numerically calculated the error-propagation factor induced by the inversion matrices, $\prod_k \|V^{(k)+}\|$. One can see that \jetext{the inversion-error propagation factor} is extremely large for the square design ($\alpha=1$, $\approx1.5\times10^{39}$), explaining the catastrophic failure observed in Fig.~\ref{fig:noisy} (a). This factor falls steeply with oversampling $\approx3.5\times10^{5}$ at $\alpha=2$, ${\approx}40$ at $\alpha=3$, and finally reaches $8.487\times 10^{-1} < 1$ by $\alpha=5$. 

These results are interesting because in Theorem~\ref{thm:error_supression}, we showed that the coefficient approximation error is guaranteed to be suppressed when the oversampling is sufficiently large, $\alpha \propto \varepsilon^{-2}\,(\log(2D+1)+\log L)$. But only $\alpha=2,3$ with shot budget $S=10^4$ was enough for comparing to noiseless baseline. Also, it is worth noting that even though $\prod_k \|V^{(k)+}\|$ is large (order of $10^5$) for $\alpha=2$, the relative error for functions behaves well, indicating our bound is loose. In conclusion, we observe that the full FTCI algorithm for noisy quantum functions is practical \jetext{with a moderate oversampling factor}.

\section{\jetext{Discussion and outlook}}
Quantum processes controlled by classical parameters naturally define quantum functions once a classical output is extracted. Such functions encompass applications from quantum simulation to quantum machine learning. Understanding when they can be learned and represented efficiently clarifies when quantum advantage might arise and has practical value: evaluating a learned classical representation can be far less costly than repeatedly querying quantum hardware.

A key structural feature we have exploited is that quantum functions can be expressed as finite linear combinations of basis functions. In many settings, these basis functions are determined by how the experimenter controls the process, so learning reduces to identifying their coefficients. Their number, however, typically grows exponentially with the input dimension, making naive regression intractable. One response is to retain a manageable subset of components and identify regimes in which the resulting approximation is accurate. Random Fourier features, truncated trigonometric kernels (low-degree approximations), and Pauli-truncation methods follow this approach, with efficiency guarantees that rely on sparsity or concentration of the coefficient vector or related learning objects.

In this work, we have studied a different notion of efficient representability: low bond dimension in a tensor-network representation of the quantum function. This concerns the function's coefficient tensor rather than the quantum states of the underlying process. The class of coefficients admitting small-bond-dimension \emph{matrix product state} (MPS) representations includes sparse coefficient tensors but extends beyond them. \emph{Tensor cross interpolation} (TCI) provides a way to learn such structure from queries. Building on it, we have introduced \emph{function TCI} 
(FTCI) as a learning primitive for functions represented in tensor-product bases.

Our central observation has been that every quantum function in our setting can be written as a contraction of a coefficient tensor with a tensor product of local feature vectors. FTCI uses this structure to target a compressed MPS representation of the coefficient tensor directly. The resulting criterion for efficient representation is the tensor's compressibility, which need not take the form of coefficient sparsity.

\jetext{This also highlights a rather different mindset for learning quantum processes. The aim is not to reconstruct or simulate a quantum-state trajectory and subsequently calculate observables from it. Instead, we learn the dependence of the observable itself on the classical parameters. The complexity of these two tasks need not coincide: Even a quantum evolution that is hard to simulate may give rise to a function with a compact classical representation. This separation between the complexity of the underlying process and that of its measured input--output behavior is at the heart of our approach.}

A remaining challenge is to connect this compressibility criterion to properties of the underlying quantum process. Beyond special cases, the relationship between the MPS entanglement of the coefficient tensor and circuit or Hamiltonian structure is nontrivial; efficiency can be difficult to certify from a process description alone. An important open question is to identify physically motivated families of processes whose coefficient tensors provably admit accurate, low-bond-dimension approximations. Low-bond-dimension MPS approximations have 
long been central to quantum many-body physics. Our results have shown 
show the same structural idea can be applied to functions of classical control parameters.

\jetext{Perhaps the most important take-home message} of this work is that our definition of quantum functions covers settings beyond the examples studied here, and FTCI itself applies to functions with suitable tensor-product feature maps whenever their coefficient tensors are compressible. More broadly, we advocate treating the dependence of measured outputs on control or variational parameters as the object to be learned directly. This perspective offers a way to study families of quantum processes through their input-output behavior and may prove useful well beyond the particular algorithm and examples developed here.

\section*{Code availability}
\edit{The code and data underlying the numerical results are available at \url{https://github.com/wookshin26/ftci}. The repository also contains \textit{FTCI Studio}, an interactive web application that runs FTCI on user-defined functions and quantum circuits and can be launched in the browser with or without local installation.}

\section*{Acknowledgements}

SS has been supported by the faculty research fund of Sejong University in 
2026, IITP (RS-2025-25464252), NRF (Nos.\ RS-2025-25464492, RS-2023-NR057243) funded by the Ministry of Science and ICT (MSIT), Korea. RS thanks the Alexander von Humboldt foundation for their support, under the German Research Chair program at the African Institutes for Mathematical Sciences. Part of this work was carried out while JE and SS visited the African Institute for Mathematical Sciences (AIMS), Cape Town, for the ``1st AIMS Workshop on the Theory of Quantum Learning Algorithms'' (2025). JE has been supported by the BMFTR (MuniQC-Atoms, QuSol, PraktiQOM), the Clusters of Excellence (ML4Q, MATH+), 
the QuantERA (SDPCode), the Munich Quantum Valley, Berlin Quantum, the Quantum Flagship (Millenion, Pasquans2), the DFG (CRC 183, SPP 2514), and the European Research Council (DebuQC).

\newpage

\clearpage
\appendix
\begin{widetext}

\section{Proof of Proposition}\label{app:proof_of_prop1}
Here, we prove Proposition~\ref{prop:TN_repre} of the main text. \jetext{For the proof, it is more convenient} to use the vectorization of quantum channels. We denote vectorized matrix $A\in \mathbb{C}^{m\times n}$ as
\begin{equation}
    |A\rangle\!\rangle \in \mathbb{C}^{mn}.
\end{equation} 
Then we have 
\begin{equation}
    \Tr{A^{\dagger} B} := \langle\!\langle A|B\rangle\!\rangle.
\end{equation} For quantum channel $C_j$ with Kraus operators $\{K_{j,i}\}_i$, in vectorized notation \jetext{their action becomes matrix multiplication} as
\begin{align}
    C_j(\rho) &= \sum_{i}K_{j,i}\rho K^{\dagger}_{j,i} \nonumber\\
    &= \left(\sum_{i}K^{*}_{j,i}\otimes K_{j,i}\right)|\rho\rangle\!\rangle \nonumber\\
    &:= \mathcal{K}_j|\rho\rangle\!\rangle .
\end{align} For encoding unitary channel $e^{-iG^{(k)}\phi^{(k)}(x)}$, we denote
\begin{equation}
    \mathcal{V}_k(x):=e^{iG^{(k)}\phi^{(k)}(x)}\otimes e^{-iG^{(k)}\phi^{(k)}(x)}. 
\end{equation}
Now using these notations, our quantum function becomes
\begin{align}
    f_Q(x;O,\mathcal{P,G}) &=\Tr{\mathcal{C}_{\mathcal{G,P}}(x)[\rho] O} \nonumber\\
    &=  \Tr\left[\prod_{k=1}^L \left(C_k\circ e^{-iG^{(k)}\phi^{(k)}(x)}\right)\circ C_0[\rho]O\right] \nonumber\\ 
    \label{eq:f_Qvectorized}
    &=\langle\!\langle O|\prod_{k=1}^L\left( \mathcal{K}_k\mathcal{V}_k(x)\right)\mathcal{K}_0 |\rho\rangle\!\rangle.
\end{align} 
Note that as long as the encoding generators $G^{(k)} \in \mathcal{G}$ s do not depend on the input, we can always diagonalize and absorb diagonal gates to non-data dependent quantum channels. Therefore, we assume that encoding generators are
\begin{equation}
    G^{(k)} = \operatorname{diag}\left[\lambda^{(k)}_1,\lambda^{(k)}_2,\ldots,\lambda^{(k)}_{m^{(k)}}\right].
\end{equation}
With diagonal generators, we see that $\mathcal{V}(x)$s are diagonal too, with elements of
\begin{align}
    \{e^{-i~\omega^{(k)}_{l,m}\phi^{(k)}(x)}\}_{l,m},
\end{align} where $l,m \in \{1,\ldots,m^{(k)}\}$, and we denote $\omega^{(k)}_{l,m} := \lambda^{(k)}_l-\lambda^{(k)}_m$. \jetext{We make the following observations:}
\begin{itemize}
    \item Among $(m^{(k)})^2$ combinations of $\lambda^{(k)}_l-\lambda^{(k)}_m$, there exist $m^{(k)}$ combinations giving us the same value $0$, when $l = m$. Moreover, more redundancy can exist depending on the eigenvalue set we have.  
    \item The maximum number of distinct $\omega^{(k)}_{l,k}$s is $2\begin{pmatrix}
    m^{(k)}\\2
\end{pmatrix} +1$
    \item $\omega_{l,m} = -\omega_{m,l}$.
\end{itemize}

Now we \jetext{define the vectors $\ket{\mathcal{V}(x)}$ to contain the diagonal components of the matrices $\mathcal{V}^{(k)}(x)$}. Then matrix multiplication in Eq.~\eqref{eq:f_Qvectorized} becomes
\begin{align}
    &\sum_{i_1,\ldots,i_L}\mathcal{K}^{(L)}_{i_{L+1},i_{L}}\mathcal{V}^{(L)}(x)_{i_{L},i_{L}}\mathcal{K}^{(L-1)}_{i_{L},i_{L-1}}\mathcal{V}^{(L-1)}(x)_{i_{L-1},i_{L-1}}\cdots\mathcal{V}^{(1)}_{i_1,i_1}\mathcal{K}^{(0)}_{i_1,i_0} \nonumber\\
    &= \sum_{i_1,\ldots,i_L}\mathcal{K}^{(L)}_{i_{L+1},i_{L}}\mathcal{K}^{(L-1)}_{i_{L},i_{L-1}}\cdots \mathcal{K}^{(0)}_{i_1,i_0}\ket{\mathcal{V}^{(L)}(x)}_{i_L}\ket{\mathcal{V}^{(L-1)}(x)}_{i_{L-1}}\cdots\ket{\mathcal{V}^{(1)}(x)}_{i_1}. \nonumber\\
\end{align}
Next, we take singular value decompositions for all $\mathcal{K}^{(k)} = U^{(k)}\Sigma^{(k)} V^{(k)\dagger}$, and contract those matrices 
to generate 3-leg tensors,
\begin{align}\label{eq:MPSmatrices}
    C'^{(k)}_{i_{k+1}i_ki_{k-1}} := V^{(k)\dagger}_{i_{k+1},i_k}\left(U^{(k-1)}\Sigma^{(k-1)}\right)_{i_ki_{k-1}}. 
\end{align}
As a result, we get a tensor network representation of $f_Q(x; O,  \mathcal{P}, \mathcal{G})$,
\begin{align}
    f_Q(x; O, \mathcal{P}, \mathcal{G}) = \mathbf{C'} \cdot \bigotimes_{k=1}^L\ket{\mathcal{V}^{(k)}(x)},
\end{align} where $\mathbf{C'}$ is the \emph{matrix product state}  (MPS) with tensor blocks composed of $C'$, as we define in Eq.~\eqref{eq:MPSmatrices}.

As a final touch, we make elements in vectors $\ket{\mathcal{V}^{(k)}(x)}$ be real values and remove redundant components. This can be done by adjusting $C'^{(k)}$s \jetext{in the following ways}. First, summing matrices
\begin{equation}
    C'^{(k)}_{i_{k+1},j_k,i_{k-1}} \leftarrow \sum_{i_k\in R_{j_k}}C'^{(k)}_{i_{k+1},i_k,i_{k-1}},\quad\quad \text{where}~~\forall (i_k,i'_k) \in R_{j_k}\times R_{j_k}, ~\ket{\mathcal{V}^{(k)}(x)}_{i_k}=\ket{\mathcal{V}^{(k)}(x)}_{i'_k}. 
\end{equation} Here $R_{j_k}$ is the set of indices where $i_k$'th elements in $\ket{\mathcal{V}^{(k)}(x)}$ are all the same. We denote $2D^{(k)}+1$ as the cardinality of $R_{j_k}$, in other words, the dimension of the index $j_k$ is $2D^{(k)}+1$. We also adjust $\ket{\mathcal{V}^{(k)}(x)}$ to have $2D^{(k)}+1$ dimension by removing redundant elements.
Second, we construct the basis-changing matrices 
\begin{equation}
    T^{(k)} = \frac{1}{\sqrt{D^{(k)}+1}}(1)\bigoplus_{j_k}\frac{1}{2}\begin{pmatrix}
    1&1\\-i&i
    \end{pmatrix},
\end{equation}\jetext{ and apply} to $\ket{\mathcal{V}^{(k)}(x)}_{j_k}$, and inverse of it to $C'^{(k)}$ with proper permutation $P$. We define
\begin{align}
    \mathbf{\Phi}^{(k)}(x) &:= T^{(k)}P\ket{\mathcal{V}(x)}= \sqrt{\frac{1}{D^{(k)}+1}}\begin{pmatrix}
        1\\ \cos\omega^{(k)}_1\phi^{(k)}(x) \\\sin\omega^{(k)}_1\phi^{(k)}(x) \\ \vdots \\\cos\omega^{(k)}_{D^{(k)}}\phi^{(k)}(x)\\ \sin\omega^{(k)}_{D^{(k)}}\phi^{(k)}(x)
    \end{pmatrix},
\end{align} where $\{w^{(k)}_{i}\}_{i=1}^{(D_k-1)/2}$ is given by the unique elements in the set $\{\lambda^{(k)}_l-\lambda^{(k)}_m\}_{l,m \in [m^{(k)}]}$. And 
\begin{align}
    C^{(k)}_{i_{k+1}i_ki_{k-1}} :=\sum_{j_k} C'^{(k)}_{i_{k+1}j_ki_{k-1}}\left(T^{(k)}P\right)^{-1}_{j_ki_k}.
\end{align} Now we have a tensor network representation of the quantum function
\begin{equation}
   f_Q(x; O, \mathcal{G}, \mathcal{P})= \mathbf{C}(O,\{C_i\}_i) \cdot \bigotimes_{k=1}^L \mathbf{\Phi}^{(k)}(\mathcal{G},\mathcal{P})(x),
\end{equation} 
where $\mathbf{C}$ is the MPS constructed from \jetext{the blocks $C^{(k)}$}\jetext{.}
This derivation shows that every quantum function using data-dependent unitaries in the form of \edit{Eq.~\eqref{eq:channel_structure}} is an 
inner product between $L$ tensor-product basis 
function vectors and coefficient tensor $\mathbf{C}$. 

\section{Noisy TCI}\label{app:Noisy}

This appendix recalls the noisy TCI estimate used to assess errors in the sampled function tensor.
We do so in several steps.

\subsection{Error scaling of noisy tensor cross interpolation}

We state the relevant reconstruction bound before examining its dependence on noise and tensor length. Let us denote the true, noiseless tensor as $\mathbf{T}$, and the noisy tensor as $\tilde{\mathbf{T}} := \mathbf{T} +\mathcal{E}$. Here, $\mathcal{E}$ is the (measurement) noise tensor, which has the same dimensions as $\mathbf{T}$ but contains non-zero elements exclusively at the queried indices. Because noise can drastically alter the condition number and rank of the queried submatrices, standard exact interpolation is unstable. However, by employing a truncated pseudoinverse, 
\emph{tensor cross interpolation} (TCI) 
provides a stable approximation $\hat{\mathbf{T}}$.

Let us define the total effective error bound as
\begin{equation}
    \bar{\varepsilon} := \varepsilon + \norm{\mathcal{E}}_F
\end{equation}
where $\varepsilon$ is the inherent approximation error of the true tensor $\mathbf{T}$ to an 
exact \emph{tensor train} or \emph{matrix product state} (MPS) format of rank $r$.
Following the analysis \jetext{in Ref.~}\cite{qin2022error}, we have the following error bound for the reconstructed tensor.

\begin{theorem}[Error bound of noisy TCI (Thm.\ 4 
of \je{Ref.}~\cite{qin2022error}) ]
Let $\mathbf{T}^{(k)} = W^{(k)}\Sigma^{(k)} {V^{(k)}}^\top$ be the compact SVD of the $k$-th unfolding matrix of the best rank-$r_k$ approximation of $\mathbf{T}$. For a spectral norm $\norm{\cdot}$, define the condition parameters
\begin{itemize}
    \item $a = \max_k \left\{ \norm{\left(W^{(k)}[I^{(k)}, :]\right)^\dagger}, \norm{\left(V^{(k)}[:, J^{(k)}]\right)^\dagger} \right\},$ 
    \item $\tilde{c} = \max_k \norm{\left(\tilde{\mathbf{T}}^{(k)}[I^{(k)}, J^{(k)}]\right)^\dagger}$.
\end{itemize}

Then, using appropriate thresholding parameters for the truncated pseudoinverse, the reconstructed tensor $\hat{\mathbf{T}}$ satisfies
\begin{equation} \label{eq:noisyerrorbound_full}
    \norm{\mathbf{T} - \hat{\mathbf{T}}}_F \le (a^2r + a^2\tilde{c}r\bar{\varepsilon} + a^2\tilde{c}^2\bar{\varepsilon}^2)^{\lceil \log_2 n \rceil - 1} \cdot (a^2\bar{\varepsilon} + a^2\tilde{c}\bar{\varepsilon}^2 + a^2\tilde{c}^2\bar{\varepsilon}^3)
.
\end{equation}
\end{theorem}

\subsection{Asymptotic error scaling}

To understand how this error scales with $n$, we must bound both the condition number $\tilde{c}$ and the Frobenius norm of the noise $\norm{\mathcal{E}}_F$. We do this in this subsection.

\begin{enumerate}
    \item \textbf{Bounding the condition number $\tilde{c}$:} Adding noise to a matrix can drastically reduce its singular values, meaning the standard triangle inequality does not apply to the pseudoinverse. However, the TCI algorithm mitigates this by applying a truncated pseudoinverse (setting singular values below a threshold $\tau_k$ to zero). With an appropriately chosen threshold, $\tilde{c}$ remains well-conditioned and is expected to be closely bounded by the noiseless condition parameter $c$.
    
    \item \textbf{Bounding the noise norm $\norm{\mathcal{E}}_F$:} The TCI algorithm is highly efficient, querying only $O(n r^2)$ elements across the entire tensor network. If the statistical measurement noise on each queried expectation value has a magnitude of $\sim \delta$, the Frobenius norm of the sparse noise tensor is given by:
    \begin{equation}
        \norm{\mathcal{E}}_F = O(\delta \sqrt{n r^2}) = O(n^{1/2} r \delta)
    \end{equation}
\end{enumerate}
Assuming the true tensor can be exactly represented by an MPS of rank $r$ ($\varepsilon = 0$), we have $\bar{\varepsilon} = O(n^{1/2} r \delta)$. Substituting this into the error bound and taking the leading-order terms (assuming $\delta$ is small), the dominant noise contribution is driven by the linear factor $a^2\bar{\varepsilon}$ amplified by the leading term of the exponent factor
\begin{equation}
    \norm{\mathbf{T} - \hat{\mathbf{T}}}_F \le O \left( (a^2r)^{\lceil \log_2 n \rceil - 1}  (a^2 n^{1/2} r \delta) \right).
\end{equation}
Therefore, the additive error introduced by the noisy measurements scales polynomially with the subsystem rank $r$ and noise $\delta$, and grows only \textit{quasi-polynomially} with respect to the tensor size $n$. This demonstrates that statistical errors encountered in quantum measurements do not proliferate exponentially when performing tensor cross approximation\jetext{.}

\section{Proof of Theorem~\ref{thm:error_supression}}\label{app:thm1_proof} 
In \jetext{function TCI}, one first uses TCI to approximate the function tensor $\mathbf{F}$, and \jetext{then applies the inversion matrices $V^{(k)+}$} that depend on the basis functions $\Phik(x)$ and auxiliary points $\{a^j_k\}_j$ we choose. Applying inversion \jetext{introduces an additional multiplicative factor in the TCI error,} which is upper bounded by the spectral norm of \begin{equation}
    \bigotimes_{k=1}(V^{(k)}(\Phik, \{a_j^{(k)}\}))^+.
    \end{equation}
Here, we show that for \jetext{the matrices $V^{(k)}$} constructed by trigonometric functions---which is the case for quantum functions---randomly chosen points 
\begin{equation}
\{a_j^{(k)}\}^{M^{(k)}}_{j=1} 
\end{equation}decrease the spectral norm as $\sim 1/\sqrt{M^{(k)}}$ with high probability.
For ease of notation, we drop the \edit{upper} index $(k)$, which was for identifying different basis functions (frequencies), as this result applies for all $k$s. Now
\begin{align}
     V &=\sqrt{\frac{1}{D+1}}\left[\begin{pmatrix}
        1 \\ \cos\omega_1a_1 \\ \sin \omega_1 a_1 \\ \vdots \\ \cos\omega_{D}a_1 \\ \sin \omega_{D} a_1
    \end{pmatrix} ,\cdots, \begin{pmatrix}
        1 \\ \cos\omega_1a_{M} \\ \sin \omega_1 a_{M} \\ \vdots \\ \cos\omega_{D}a_{M} \\ \sin \omega_{D} a_{M}
    \end{pmatrix}  \right] \nonumber\\ 
    &:= \left[ v(a_1), v(a_2),\cdots, v(a_M) \right] .
\end{align}
One can note that this is a design matrix with feature vector $v(\cdot)$s, and all vectors are normalized.

To prove Theorem~\ref{thm:error_supression}, first we have a concentration result.

\begin{lemma}[Eigenvalue concentration (Ref.~\cite{tropp2012user}, Remark 5.3.\ 
matrix Chernoff with $R=1$)]\label{lem:concentration}
Let $a_1,\ldots,a_M\stackrel{\text{i.i.d.}}{\sim}\mathrm{Unif}[0,l]$ and $V\in\mathbb{R}^{N\times M}$ be the design matrix with columns $v(a_i)$, with $\|v(a_i)\|_2^2=1$.
Set $G=\frac{1}{M}VV^\top$,and $\Sigma = \underset{x}{\mathbb{E}} \left[v(x)v(x)^{\top}\right].$

Then for any $\varepsilon\in(0,1)$,
\begin{equation}
    \mathbb{P}\left\{~\lambda_{\min}(G)\ \geq (1-\varepsilon)\lambda_{\min}(\Sigma)\right\}
 \geq\ 1 - N\exp\Big(-\lambda_{min}(\Sigma)M\varepsilon^2/2\Big)\je{.}
\end{equation}
\end{lemma}
As a corollary \je{of this statement}, 
with the same probability,
\je{we find}
\begin{equation}
    \sigma_{\min}(V)\ =\ \sqrt{M\,\lambda_{\min}(G)}\ \ge\ \sqrt{M(1-\varepsilon)\lambda_{\min}(\Sigma)}.
\end{equation}
 We use the following.

\begin{theorem}[Ingham–Montgomery–Vaughan inequality with explicit constants]\label{thm:IMV}
Let $\{\omega_l\}\subset\mathbb{R}$ be \emph{uniformly separated}: there exists $\gamma>0$ with
$\inf_{l\neq m}|\omega_l-\omega_m|\ge\gamma$.
Then for every $P>0$ and every finitely supported $(a_\ell)$,
\begin{equation}\label{eq:IMV}
\big(P-\tfrac{2\pi}{\gamma}\big)\sum_\ell |a_\ell|^2
\ \le\ \int_0^P\Big|\sum_\ell a_\ell e^{i\omega_l t}\Big|^2 dt
\ \le\ \big(P+\tfrac{2\pi}{\gamma}\big)\sum_l |a_l|^2.
\end{equation}

\end{theorem}

\begin{proof}[Proof]
Write
\begin{equation}
\int_0^P\Big|\sum_\ell a_\ell e^{i\omega_l t}\Big|^2 dt
= P\sum_\ell |a_\ell|^2
+ \sum_{\ell\ne m} a_\ell \overline{a_m}\,\frac{e^{i(\omega_l-\omega_m)P}-1}{i(\omega_l-\omega_m)}.
\end{equation}
Set $F(x)=\tfrac{e^{ixP}-1}{ix}=e^{ixP/2}\cdot\tfrac{2\sin(xP/2)}{x}$. Using the Montgomery–Vaughan quadratic form bound~\cite{montgomery1974hilbert}
\begin{equation}
\Big|\sum_{\ell\ne m}\frac{a_\ell \overline{a_m}}{\omega_l-\omega_m}\Big|\ \le\ \frac{\pi}{\gamma}\sum_\ell |a_\ell|^2,
\end{equation}
and $|ie^{i(\cdot)}\sin(\cdot)|\le 1$, yields \eqref{eq:IMV}.  \qedhere
\end{proof}
Consequently, if $P>2\pi/\gamma$, the complex exponentials $\{e^{i\omega_l t}\}$ form a Riesz sequence in $L^2(0,P)$
with Riesz bounds 
\begin{equation}
A'_{P,\gamma}=P-2\pi/\gamma,\quad B'_{P,\gamma}=P+2\pi/\gamma
\end{equation} \jetext{These} depend only on $P$ and the gap $\gamma$ (and \emph{not} on how many frequencies are used).

\begin{lemma}[Riesz bounds for the real dictionary]\label{lem:real-riesz}
Assume the \emph{full} frequency set $\Omega=\{0,\pm\omega_1,\ldots,\pm\omega_D\}$ is $\gamma$-separated
\begin{equation}
\gamma \le \min\Big\{\,\min_j|\omega_j|,\ \min_{j\neq k}|\omega_j-\omega_k|~\,\Big\}.
\end{equation}
If $P>2\pi/\gamma$, then $\mathcal{F}_{\mathrm{real}}$ is a Riesz sequence in $L^2(0,P)$ with bounds
\begin{equation}
A_{P,\gamma}\,\|b\|_2^2 \ \le\ \int_0^P\Big|\,b_0+\sum_{j=1}^D\big(b_{j,c}\cos(\omega_j t)+b_{j,s}\sin(\omega_j t)\big)\,\Big|^2 dt
\ \le\ B_{P,\gamma}\,\|b\|_2^2,
\end{equation}
where $A_{P,\gamma},B_{P,\gamma}$ depend only on $P$ and $\gamma$ and can be taken as
\begin{equation}
    A_{P,\gamma} = \frac{P-2\pi/\gamma}{2},\qquad
B_{P,\gamma} = P+2\pi/\gamma. 
\end{equation} 
\end{lemma}

\begin{proof}
From Theorem~\ref{thm:IMV} we have Eq.~\eqref{eq:IMV} with the complex system on $\Omega$, having $A'_P = P-2\pi/\gamma,~B'_P = P +2\pi/\gamma$. 
Now we change the bases from complex to real. Let $\Omega=\{0\}\cup\{+\omega_j,-\omega_j\}_{j=1}^D$. \jetext{Between the complex basis system} $\mathcal{F}_{\mathrm{cx}}=\{e^{i\omega t}:\omega\in\Omega\}$, \jetext{and the real basis system} $\mathcal{F}_{\mathrm{real}}=\{\,1,\ \cos(\omega_j t),\ \sin(\omega_j t)~|~ j\in[D]~\}\subset L^2(0,P),$ there is a fixed linear change of variables
\begin{equation}
    \cos(\omega t)=\tfrac12(e^{i\omega t}+e^{-i\omega t}),\quad
\sin(\omega t)=\tfrac{1}{2i}(e^{i\omega t}-e^{-i\omega t}).
\end{equation}
This maps the real coefficients $b$ to complex coefficients blockwise via
\begin{equation}
a_{j,+}=\tfrac12(b_{j,c}-i b_{j,s}),\quad a_{j,-}=\tfrac12(b_{j,c}+i b_{j,s}),\quad a_0=b_0,
\end{equation}
so that $b_{j,c}\cos(\omega_j t)+b_{j,s}\sin(\omega_j t)=a_{j,+}e^{i\omega_j t}+a_{j,-}e^{-i\omega_j t}$.
On each $(\cos,\sin)$ block, $\|(a_{j,+},a_{j,-})\|_2=\|(b_{j,c},b_{j,s})\|_2/\sqrt{2}$, hence globally
$\|a\|_2 \le \|b\|_2 \le \sqrt{2}\,\|a\|_2$,
where $a$ and $b$ are coefficient vectors in the complex and real bases, respectively.
Finally, we transfer Eq.~\eqref{eq:IMV} to the real system via the bounded change-of-basis $a=Tb$, yielding 
\begin{equation}
    A_{P,\gamma} = A'_P/2\qquad B_{P,\gamma} = B'_P.
\end{equation}
\end{proof}

Now we are ready to prove the theorem.

\begin{theorem}[Spectral norm bound for one design matrix with trigonometric functions]\label{thm:spectral_bound_for_one}
Fix $D\ge1$, and set
\begin{equation}
    v(t)=\frac{1}{\sqrt{D+1}}\big(1,~ \cos(\omega_1 t),~ \sin(\omega_1 t),~ \ldots,~ \cos(\omega_D t),~ \sin(\omega_D t)\big)^\top\in\mathbb{R}^{N}.
\end{equation} A design matrix $V\in \mathbb{R}^{(2D+1)\times M}$ is constructed by stacking $M$ $v(a^i)$ vectors with $a^1,\ldots,a^{M} \stackrel{\text{i.i.d.}}{\sim}\mathrm{Unif}[0,P]$. Then,

\begin{equation}
    \|V^{+}\|_2\ =\frac{1}{\sigma_{\min}(V)}
\le\sqrt{\frac{C(P,\gamma)}{(1-\varepsilon)}}\ \sqrt{\frac{D+1}{M}},
\end{equation} with probability at least $1-N\exp\left[-\frac{M\varepsilon^2}{2C(P,\gamma)(D+1)}\right]$.
The spectral gap $\gamma$ is defined as,
\begin{equation}
    \gamma \le \min\Big\{\,\min_j|\omega_j|,\ \min_{j\neq k}|\omega_j-\omega_k|~\,\Big\},
\end{equation} $\varepsilon \in (0,1)$,and $C(P,\gamma) := \frac{2P}{P-2\pi/\gamma}$

\end{theorem}

\begin{proof}
    Let us denote $\Sigma:=\mathbb{E}\big[v(t)v(t)^\top\big]\in\mathbb{R}^{N\times N}.$ Note that 
    \begin{align}
        \lambda_{min}(\Sigma) &= \min_{\|b\|_2=1}~b^\top\Sigma b =\min_{\|b\|_2=1}~\avg{t}{\left| v(t) \cdot b \right|^2}
    \end{align}

    Let $t\sim\mathrm{Unif}[0,P]$. Then for every $b\in\mathbb{R}^{N}$,
    \begin{equation}\label{eq:sigma-quad}
        \avg{t}{\left|v(t) \cdot b \right|^2} =\frac{1}{P}\int_0^P(v(t)\cdot b)^2\ dt = \frac{1}{P(D+1)}\int_0^P \Big|\,b_0+\sum_{j=1}^D\big(b_{j,c}\cos(\omega_j t)+b_{j,s}\sin(\omega_j t)\big)\,\Big|^2 dt.
    \end{equation} 
    Since $\|v(t)\|_2\equiv 1$, \eqref{eq:sigma-quad} follows by expanding $v(t)\cdot b$. Consequently, using Lemma~\ref{lem:real-riesz} under the separation $\gamma$ and the length condition $P>2\pi/\gamma$,
    \begin{equation}
        \frac{A_{P,\gamma}}{P(D+1)}\,\|b\|_2^2 \ \le\ b^\top\Sigma\,b \ \le\ \frac{B_{P,\gamma}}{P(D+1)}\,\|b\|_2^2,
    \end{equation}
    i.e.,
    \begin{equation}
        \frac{A_{P,\gamma}}{P(D+1)} \le\lambda_{\min}(\Sigma)\ ,\qquad
    \lambda_{\max}(\Sigma)\ \le\ \frac{B_{P,\gamma}}{P(D+1)}.
    \end{equation}
    Now it follows from Lemma~\ref{lem:concentration},
    \begin{align}
          \|V^{+}\|_2 = \frac{1}{\sigma_{min}(V)}&\leq \frac{1}{\sqrt{M(1 -\varepsilon)\lambda_{min}(\Sigma)}} \nonumber\\
          &\leq\sqrt{\frac{P(D+1)}{M(1-\varepsilon)A_{P,\gamma} }},
    \end{align}
   with probability at least
   \begin{equation}
       1-(2D+1)\exp\left[-\frac{A_{P,\gamma}M\varepsilon^2}{2P(D+1)}\right]
       .
   \end{equation}
\end{proof}
The above theorem tells us that with a probability that approaches one exponentially fast with $M$, the spectral norm scales as $O\left(\sqrt{\frac{D}{M}}\right)$. Now we have a corollary for tensor products.

\begin{corollary}[Product bound for local design inverses]

Let \begin{align}
    V^{(k)}(a^{(k)}) = \sqrt{\frac{1}{D^{(k)}+1}}\left[\begin{pmatrix}
        1 \\ \cos\omega^{(k)}_1a^{(k)}_1 \\ \sin \omega^{(k)}_1 a_1^{(k)} \\ \vdots \\ \cos\omega^{(k)}_{D^{(k)}}a_1^{(k)} \\ \sin \omega^{(k)}_{D^{(k)}} a_1^{(k)}
    \end{pmatrix} ,\cdots, \begin{pmatrix}
        1 \\ \cos\omega^{(k)}_1a_{M^{(k)}}^{(k)} \\ \sin \omega^{(k)}_1 a_{M^{(k)}}^{(k)} \\ \vdots \\ \cos\omega^{(k)}_{D^{(k)}}a_{M^{(k)}}^{(k)} \\ \sin \omega^{(k)}_{D^{(k)}} a_{M^{(k)}}^{(k)}
    \end{pmatrix}  \right],
\end{align} where 
\begin{align}
    a^{(k)}_1,\ldots,a^{(k)}_{M^{(k)}} \stackrel{\text{i.i.d.}}{\sim}\mathrm{Unif}[0,P^{(k)}]
\end{align} for all $k \in [L]$, and the spectral gap \begin{equation}
    \gamma^{(k)} \le \min\Big\{\,\min_j|\omega^{(k)}_j|,\ \min_{i\neq j}|\omega^{(k)}_i-\omega^{(k)}_j|~\Big\}.
\end{equation} Then, we have
\begin{align}
    \norm{\bigotimes_{k=1}^LV^{(k)+}(a_k)} \leq \prod_{k=1}^L\left(
    \sqrt{ \frac{C(P^{(k)},\gamma^{(k)})}{(1-\varepsilon)}}    \sqrt{ \frac{D^{(k)}+1}{M^{(k)}} }
    \right),
\end{align}with probability at least
\begin{equation}
    1-\sum_{k=1}^L(2D^{(k)}+1)\exp\left[-\frac{M^{(k)}\varepsilon^2}{2C(P^{(k)},\gamma^{(k)})(D^{(k)}+1)}\right].
\end{equation}
\end{corollary}

This follows from the multiplicity of the spectral norm of the tensor product, and the union bound for joint events.
Now let \begin{equation}
    M = \underset{k}{\min}M^{(k)},~D = \underset{k}{\max}D^{(k)},~ C= \underset{k}{\max}C(P^{(k)},\gamma^{(k)}).
\end{equation} 

Using this, we get the looser but clearer bound, written as the following corollary.

\begin{corollary}[Uniform inversion bound]\label{eq:coro2} Given the 
\jetext{same setting as in Corollary 1,}
    \begin{align} \mathbb{P}\left(\norm{\bigotimes_{k=1}^LV^{(k)+}(a_k)} \leq \left(\frac{C}{(1-\varepsilon)}\frac{D+1}{M}\right)^{L/2}\right) \ge 1- \delta,
\end{align} with 
\begin{equation}
    M \geq \frac{2C(D+1)}{\varepsilon^2}\left(\log(2D+1) + \log(L)+\log\left(\frac{1}{\delta}\right) \right).
\end{equation}
\end{corollary}

Note that $\frac{C}{(1-\varepsilon)}\frac{D+1}{M} \leq\frac{\varepsilon^2}{2B(1-\varepsilon)}$ for some $B\geq 1$, and \begin{equation}
    \frac{\varepsilon^2}{2B(1-\varepsilon)} \leq \frac{\varepsilon^2}{2(1-\varepsilon)} \leq 1,\qquad \varepsilon\in[0,\sqrt{3}-1].
\end{equation} Therefore, we can make sure that our spectral norm vanishes with increasing $L$. 

\section{Spreading upper bounds the entanglement}\label{app:entanglement_bound}

Let $S^{(\alpha)}(C)$ denote \jetext{R\'enyi-}$\alpha$ entropy of a given normalized $L$-leg tensor $C$, $\norm{C}_F = 1$, as
\begin{equation}
    S^{(\alpha)}(C) := \frac{1}{1-\alpha}\log\sum_{i_1,\ldots,i_L}|C_{i_1,\ldots 
    , i_L}|^{2\alpha}.
\end{equation} \jetext{Now} $E^{(\alpha)}_{(A:B)}(C)$ denotes the entanglement entropy of $C$ with bipartition of indices $(A:B)$. Then we have the following.

\begin{theorem}[Coefficient spreading bounds bipartite entanglement]
    Let $C$ be \jetext{an $L$-leg tensor} with normalization $\norm{C}_F = 1$. \jetext{For every bipartition} of indices $(A:B)$ \jetext{and every $\alpha\geq 0$ (using the continuous extension at $\alpha=1$)},
    \begin{equation}
        \jetext{E^{(\alpha)}_{(A:B)}(C) \leq S^{(\alpha)}(C)}.
    \end{equation}
\end{theorem}

\begin{proof}
    Let $M$ be a matrix that is constructed by the tensor $C$, interpreting $i_k\in A$ as row indices and $i_k\in B$ as columns. Now we have $\norm{M}_F = 1$, and
    \begin{equation}
            MM^{\dag} = \begin{pmatrix}
                &\sum_{i}\abs{M_{1,i}}^2 &\cdots&\cdots\\
                &\vdots &\sum_{i}\abs{M_{2,i}}^2&\cdots\\
                &\vdots&\cdots&\vdots\\
                &\vdots&\cdots&\sum_{i}\abs{M_{D,i}}^2
            \end{pmatrix}
        \end{equation} is a $D\times D$ Hermitian matrix with trace 1. By the Schur-Horn theorem, the vector of eigenvalues of $MM^{\dag}$, \jetext{which are the squares of the singular values} of $M$,
        \begin{equation}
            \vec{\sigma^2}(M) := \begin{pmatrix} \sigma_1(M)^2\\\sigma_2(M)^2\\\vdots\\\sigma_D(M)^2
            \end{pmatrix}
        \end{equation}
        majorizes the ${\rm diag}(MM^{\dag})$. 
        Now we construct 
        a $D^2$-dimensional vector 
        \begin{equation}
            \vec{\sigma^2}_{D^2}(M) := [\sigma_1^2(M), 0, \ldots, \sigma_2^2(M), 0, \ldots\sigma^2_D(M), \ldots , 0 ], 
        \end{equation} and
        \begin{equation}
            \vec{|M|^2} := [|M_{1,1}|^2~|M_{1, 2}|^2~\ldots~|M_{2, 1}|^2~|M_{2,2}|^2~\ldots~|M_{D, 1}|^2~\ldots~|M_{D, D}|^2].
        \end{equation}
        We observe that $\vec{\sigma^2}_{D^2}(M) \succeq \vec{|M|^2}$ follows from the $\vec{\sigma^2}(M) \succeq {\rm diag}(MM^{\dag}) \succeq\vec{|M|^2} $. As \jetext{R\'enyi-}$\alpha$ 
        entropy is Schur-concave, 
        \begin{equation}
        S^{(\alpha)}(\vec{\sigma^2_{D^2}}(M)) = S^{(\alpha)}(\vec{\sigma^2}(M)) \leq S^{(\alpha)}(\vec{|M|^2}).
          \end{equation}
        We conclude the proof by noting that $S^{(\alpha)}(\vec{\sigma^2}(M)) = E^{(\alpha)}_{(A:B)}(C)$ and $S^{(\alpha)}(\vec{|M|^2}) = S^{(\alpha)}(C) $. The proof does not depend on the bipartition $(A:B)$.
\end{proof}

\end{widetext}

%


\end{document}